\documentclass[aps,prb,twocolumn,superscriptaddress,floatfix,letterpaper,nofootinbib]{revtex4-2}

\usepackage[a-1b]{pdfx}

\newcommand{\hstar}{\mathop{*}} 
\def\wdg{{\mathchoice{\,{\scriptstyle\wedge}\,}{{\scriptstyle\wedge}}{{\scriptscriptstyle\wedge}}{{\scriptscriptstyle\wedge}}}} 
\definecolor{refkey}{rgb}{0,0.7,0}
\definecolor{labelkey}{rgb}{0,0.7,0}

\newcommand{\Sq}{\text{Sq}}

\newcommand{\cRep}{\mathcal{R}\mathrm{ep}}
\newcommand{\cVec}{\mathcal{V}\mathrm{ec}}

\newcommand{\cPSG}{\cP\mathrm{SG}}

\newcommand{\Hom}{\mathrm{Hom}}

\usepackage[normalem]{ulem}

\usepackage{tikz}
\usetikzlibrary{matrix,arrows,calc}
\tikzset{cd/.style={matrix of math nodes,row sep=2em,column sep=2em, text height=1.5ex, text depth=0.5ex}}
\tikzset{cdar/.style={->,auto}} 

\tikzset{triar/.style={anchor=mid,->}} 
\tikzset{tridar/.style={anchor=mid,double,double equal sign distance,-implies}}

\theoremstyle{plain}
\newtheorem{theorem}{Theorem}[section]

\theoremstyle{definition}

\newtheorem{definition}[theorem]{Definition}

\newtheorem{Example}[theorem]{Example}

\theoremstyle{remark}

\newcommand*{\Simp}[1]{\Delta^{#1}}
\newcommand*{\Horn}[2]{\Lambda^{#1}_{#2}}
\newcommand{\Sets}{\mathsf{Sets}}
\newcommand{\sSets}{\mathsf{sSets}}
\newcommand*{\id}{\textup{id}}
\DeclareMathOperator{\Kan}{Kan}     

\usepackage{braket}

\usepackage{array}
\newcolumntype{P}[1]{>{\centering\arraybackslash}p{#1}}
\newcolumntype{M}[1]{>{\centering\arraybackslash}m{#1}}
\renewcommand{\arraystretch}{1.75}
\usepackage{hhline}

\newcommand\T{\mathbb{T}}

\begin{document}

\begin{titlepage}

\title{Generalized symmetries in singularity-free nonlinear $\sigma$ models\\
and their disordered phases}

\author{Salvatore D. Pace}
\affiliation{Department of Physics, Massachusetts Institute of Technology,
Cambridge, MA 02139, USA}

\author{Chenchang Zhu} 
\affiliation{Mathematics Institute, Georg-August-University of
  G\"ottingen, G\"ottingen 37073, Germany}

\author{Agn\`es Beaudry}
\affiliation{Department of Mathematics, University of Colorado Boulder, 
Boulder, CO 80309-0395, USA }

\author{Xiao-Gang Wen} 
\affiliation{Department of Physics, Massachusetts Institute of Technology,
Cambridge, MA 02139, USA}

\date{\today}

\begin{abstract} 

We study the nonlinear $\sigma$-model in ${(d+1)}$-dimensional spacetime with
connected target space $K$ and show that, at energy scales below singular field
configurations (such as vortices), it has an emergent non-invertible higher
symmetry. The symmetry defects of the emergent symmetry are described by the
$d$-representations of a discrete $d$-group $\mathbb{G}^{(d)}$ (i.e. the
emergent symmetry is the dual of the invertible  $d$-group $\mathbb{G}^{(d)}$
symmetry).  The $d$-group $\mathbb{G}^{(d)}$ is determined such that its
classifying space $B\mathbb{G}^{(d)}$ is given by the $d$-th Postnikov stage of
$K$.  In $(2+1)$D and for finite $\mathbb{G}^{(2)}$, this symmetry is always
holo-equivalent to an invertible ${0}$-form---ordinary---symmetry with
potential 't Hooft anomaly.  The singularity-free disordered phase of the
nonlinear $\sigma$-model spontaneously breaks this symmetry, and when
$\mathbb{G}^{(d)}$ is finite, it is described by the deconfined phase of
$\mathbb{G}^{(d)}$ higher gauge theory.  We consider examples of such
disordered phases. We focus on a singularity-free $S^2$ nonlinear
$\sigma$-model in ${(3+1)}$D and show that it has an emergent non-invertible
higher symmetry. As a result, its disordered phase is described by axion
electrodynamics and has two gapless modes corresponding to a photon and a
massless axion. Notably, this non-perturbative result is different from the
results obtained using the $S^N$ and $\mathbb{C}P^{N-1}$ nonlinear $\si$-models
in the large-$N$ limit.


\end{abstract}

\maketitle

\end{titlepage}

\setcounter{tocdepth}{1} 
{\small \tableofcontents }

\section{Introduction}

Nonlinear ${\si}$-models are field theories with broad applications in physics.
For example, when an ordinary continuous symmetry ${G}$ is spontaneously broken
down to a subgroup ${H}$, the spontaneous symmetry broken phase can be
described by a nonlinear ${\si}$-model with target space ${G/H}$. As a
continuum field theory, this nonlinear ${\si}$-model describes the low-energy
fluctuations of Goldstone bosons~\cite{CWZ692247,CCW692250, WM12030609,
WH14027066}.  In quantum field theory, a quantum nonlinear ${\si}$-model is
define by path integral over continuous field configurations.
Such a quantum nonlinear ${\si}$-model is called a
singularity-free nonlinear $\sigma$-model.

Considering nonlinear ${\si}$-models on a spacetime lattice is also
interesting. In this paper, we will call a nonlinear $\si$-model on a
triangulated spacetime a \textit{discrete} nonlinear ${\si}$-model when its
target space is also triangulated~\cite{ZW180809394}.  The triangulated target
space no longer has the $G$-symmetry of the symmetric target space $G/H$.  In
this paper, we will not consider such a $G$-symmetry.

Discrete nonlinear
$\si$-models are interesting because they realize similar physics as their
continuum counterparts while also capturing strong coupling limits. In the
context of spontaneous symmetry breaking, discrete nonlinear ${\si}$-models are
useful in describing the phase transitions between the symmetry-breaking and
symmetric phases.

An important property of discrete nonlinear ${\si}$-models is the presence of
dynamical excitations corresponding to singularities in the continuous
nonlinear ${\si}$-models. Those singularities in continuous theory
are classified by the topology of
the target space (see \Rf{M1979591} for a classic review or Sec.~2 of
\Rf{P230805730} for a modern perspective.). Although there are no literal
singularities in discrete nonlinear ${\si}$-models, as there are no continuum
fields, corresponding excitations do exist, and we will still call them
``singularities.''\footnote{\label{Singularityfootnote}The name for the
singularities differs among communities and time, and they are sometimes also
called topological solitons or topological defects.  } 

In continuum nonlinear ${\si}$-models, sometimes, singularities
are regarded as nondynamical probes added in by hand (usually in field theory
context), while other times, singularities are regarded as dynamical
excitations (usually in condensed matter context).  To avoid confusion, in this
paper, we will use the term ``nonlinear ${\si}$-model'' to refer to a model
that allows dynamical singularities, whether continuous or discrete.  We will
use the term ``singularity-free nonlinear ${\si}$-model'' to refer to a model
whose singularities are all nondynamical probes. We will refer to dynamical
singularities as just singularities, while nondynamical singularities are
singularity defects. In this paper, we will work in Euclidean spacetime.

Singularities in nonlinear ${\si}$-models play an important role in the phase
transitions between the symmetry-breaking and symmetric phases. Indeed,
starting in the symmetry-broken---ordered---phase and proliferating
singularities drives a transition to the
symmetric---disordered---phase~\cite{B1971493, B1972610, KT1972L124, KT7381,
HN197841, NH197919, LD8851, KM19931911}. However, a phase transition between
the symmetry breaking and a disordered phase can occur without proliferating
singularities. Whereas proliferating singularities drive transitions to trivial
disordered phases, disordered phases without proliferated singularities are
nontrivial and can include topological orders and emergent gauge
bosons~\cite{CSc9311045, LTc9501101, OV0311222, XL10125671, GS10125669,
ZW180809394, P230805730}.  

The simplest way to prevent singularities from proliferating is using a
discrete nonlinear $\si$-model. The singularities can be removed by replacing
the maps from the triangulated spacetime to the triangulated target space by
simplicial homomorphisms.  We will call such discrete nonlinear $\si$-model a
singularity-free discrete nonlinear ${\si}$-model. We emphasize that
``singularity-free'' refers to the absence of dynamical singularities and that
singularity defects can still be included by hand to probe the theory.

We will ground our investigation using generalized symmetries of the
singularity-free discrete nonlinear ${\si}$-model, which correspond to emergent
symmetries below the singularity energy scale of a discrete nonlinear
${\si}$-model. In recent years, it has been realized that emergent symmetries
can be very general and contain so-called generalized symmetries (see
\Rfs{M220403045, CD220509545, S230518296, LWW230709215} for recent reviews).
Using generalized symmetries provides a classification scheme for the
singularity-free discrete nonlinear ${\si}$-model's disordered phases since
they are symmetry-breaking phases of generalized symmetries~\cite{P230805730}.

While codimension-1 invertible symmetry defects in spacetime generate ordinary
symmetries, generalized symmetries can be generated by codimension-${(p+1)}$
defects (${p}$-form symmetries~\cite{NOc0605316, NOc0702377, GW14125148,
WZ230312633, BS230402660, BBG230403789}) and non-invertible defects
(non-invertible symmetries~\cite{BT170402330, T171209542, CY180204445,
TW191202817, KZ200308898, KZ200514178, BS230517159, BBG230517165, S230800747,
PZh0011021, CSh0107001, FSh0204148, FSh0607247, FS09095013, KZ191213168,
I210315588, Q200509072, DR11070495, FT220907471, SS230702534,  PDL240612962}). While ordinary symmetries
are described by groups, invertible generalized symmetries are described by
higher groups~\cite{KT13094721, S150804770, CI180204790, BH180309336,
BCH221111764} and generic generalized symmetries are described by monoidal
higher categories~\cite{KZ200308898, KZ200514178, BBS220406564, BSW220805973,
BBF220805993, BBF221207393, BST221206159, DT230101259}. Just like ordinary
symmetries, these generalized symmetries can spontaneously
break~\cite{GW14125148, L180207747, HI180209512, KH190504617, QRH201002254,
HHY200715901, DKR211012611, RW211212735, KN220401924, EI221109570,
OPH230104706, W181202517}, have 't Hooft anomalies~\cite{GKK170300501,
KR180505367, P180201139,DT180210104,WW181211968,GW181211959,WW181211955,
HS181204716, WW181211967, DM190806977, CO191004962, KZ200514178, ES210615623,
KNZ230107112, ZC230401262,CY180204445, KOR200807567, CCH211101139,
CCH220409025, ACL221214605}, and characterize symmetry protected topological
phases~\cite{Y150803468, TK151102929, YDB180302369, W181202517, WW181211967,
TW190802613, TW191202817, KZ200308898, KZ200514178, JX200705002, JW200900023,
HJJ210509454, BDG211009529, Y211012861, I211012882, MF220607725, PW220703544,VBV221101376, TRV230308136, SS240401369, PLA240918113}. 

Among all these possible generalized symmetries, what are the generalized
symmetries in the singularity-free nonlinear ${\si}$-model? In this paper, we
show that a singularity-free nonlinear ${\si}$-model has a non-invertible
higher symmetry.  If singular field configurations are allowed, a nonlinear
${\si}$-model has an emergent non-invertible higher symmetry at energy scales
below the singularities.  

To describe such a  non-invertible higher symmetry, we remark that if the
symmetry charges of a symmetry are described by ${d}$-representations of the
${d}$-group ${\mathbb{G}^{(d)}}$, we say that the symmetry is described by the
${d}$-group ${\mathbb{G}^{(d)}}$.  In that case the symmetry defects are
described by the elements of the ${d}$-group ${\mathbb{G}^{(d)}}$.
On the other hand, if the symmetry defects of a symmetry are described by
${d}$-representations of the ${d}$-group ${\mathbb{G}^{(d)}}$, then the
symmetry is not described by the ${d}$-group ${\mathbb{G}^{(d)}}$.  In this
case, we say that the symmetry is described by fusion $d$-category
$d\text{-}\cRep(\mathbb{G}^{(d)})$.  

With such an understanding, the generalized symmetry, in a singularity-free
nonlinear ${\si}$-model with a connected target space $K$ in
${(d+1)}$-dimensional spacetime, is a $d\text{-}\cRep(\mathbb{G}^{(d)})$
symmetry.  Here $\mathbb{G}^{(d)}$ is determined such that its classifying
space $B\mathbb{G}^{(d)}$ is given by the $d$-th Postnikov stage of $K$, which
we realize as a simplicial set. This symmetry depends only on the topology of the field's target space, and, therefore, it cannot be broken or modified by local modifications to the action.

Using the notion of the dual symmetry introduced in \Rf{JW191213492,
KZ200514178},  the dual of ${d}$-group ${\mathbb{G}^{(d)}}$ symmetry is
$d\text{-}\cRep(\mathbb{G}^{(d)})$ symmetry.  Thus, we may also say that the
generalized symmetries in the singularity-free nonlinear ${\si}$-model is the
dual of the ${d}$-group ${\mathbb{G}^{(d)}}$ symmetry.  We remark that a
symmetry and its dual symmetry are holo-equivalent, \ie their corresponding
symmetric systems are identical when restricted with symmetric sub-Hilbert
spaces.  As a result, their phase diagrams and critical points are in
one-to-one correspondence.\footnote{Dual symmetries can also be discussed in
terms of discrete gauging. Suppose starting from a theory $\mathfrak{T}$ with
symmetry $\cS$, gauging $\cS$ produces a new theory ${\mathfrak{T}/\cS}$ with
symmetry $\cS^\vee$~\cite{Vafa:1989ih}. The dual symmetry $\cS^\vee$ is
determined by the original symmetry $\cS$ and the technical details of the
gauging procedure (for example, choices of discrete
torsion~\cite{Vafa:1986wx}). When $\cS$ is a finite symmetry, $\cS^\vee$ is
always a nontrivial finite symmetry. Importantly, in this case, there is always
a gauging procedure of $\cS^\vee$ that returns ${\mathfrak{T}/\cS}$ to
$\mathfrak{T}$. Therefore, the theory $\mathfrak{T}$ and $\mathfrak{T}/\cS$
contains the same physical information. This includes a one-to-one
correspondence of their phases and phase transitions. Thus, the phases and
phase transitions of $\mathfrak{T}$ can be inferred from those of
$\mathfrak{T}/\cS$}

Since fusion $d$-category
$d\text{-}\cRep(\mathbb{G}^{(d)})$ is a local fusion $d$-category, both
$d\text{-}\cRep(\mathbb{G}^{(d)})$ symmetry and its dual $\mathbb{G}^{(d)}$
symmetry are anomaly-free~\cite{KZ200308898, KZ200514178}. Thus, the
generalized symmetries in the singularity-free nonlinear ${\si}$-model are
anomaly-free.\footnote{Anomaly-free non-invertible higher symmetry is also
called algebraic higher symmetry~\cite{KZ200308898, KZ200514178}.} 

This paper is closely related to~\Rfs{ZW180809394, P230805730, CT230700939}.
\Rf{ZW180809394} studied the disordered phase of a singularity-free discrete
nonlinear ${\si}$-model in ${3+1}$ dimensional spacetime whose target space
${K}$ satisfied ${\pi_i(K) = \text{finite group}}$ for ${i=1,2}$ while
${\pi_i(K) = 0}$ for all other ${i}$.  It was shown that the disordered phase
is described by the deconfined phase of $d$-gauge theory of a $d$-group
$\mathbb{G}^{(d)}$, where the classifying space of $\mathbb{G}^{(d)}$ is the
target space $K$.  According to a modern understanding, this implies that the
disordered phase has an emergent symmetry, which is the dual of the
$\mathbb{G}^{(d)}$ symmetry. In other words, the symmetry defects of the
emergent symmetry are the electric charges of the $d$-gauge theory.  We remark
that, although a $\mathbb{G}^{(d)}$ symmetry is always invertible, its dual may
be non-invertible.  \Rf{P230805730} studied these non-invertible higher
symmetries in the deep IR of generic ordered phases and their symmetry breaking
and mixed 't Hooft anomalies. \Rf{CT230700939} studied them in the context of
continuum nonlinear ${\si}$-models and understood them using non-invertible
cohomology theories on the target space ${K}$ with topological quantum field
theory coefficients. Where they overlap, the results obtained in this paper
agree with those from \Rfs{P230805730, CT230700939}.

In this paper, we study discrete nonlinear $\si$-models and use simplicial
homomorphisms to implement singularity-free conditions, which give rise to
generalized symmetries. This formalism allows us to write down a discrete
nonlinear ${\si}$-model with terms that suppress its singularities. This makes
the notation of emergent generalized symmetries in generic ordered phases from
\Rf{P230805730} more precise. Furthermore, using such rigorously defined
discrete nonlinear ${\si}$-models, as well as modeling $B\mathbb{G}^{(d)}$ as a
simplicial set, we can more rigorously compute the generalized symmetry in the
singularity-free nonlinear ${\si}$-model. 

The paper is organized as follows.  In Section~\ref{dismodel}, we review
discrete nonlinear ${\si}$-models and introduce singularity-free discrete
nonlinear ${\si}$-models. In Sections~\ref{emsymm} and~\ref{piinf}, we compute
the non-invertible higher-form symmetries in singularity-free nonlinear
${\si}$-models, first for the case where ${\pi_{i\leq d}(K) = \text{finite
group}}$ and then where these homotopy groups are non-finite. In
Section~\ref{examples}, we discuss some simple examples of generalized
symmetries of singularity-free nonlinear ${\si}$-models and their disordered
phases.

One of the main results of this paper is the example presented in Section~\ref{S2Section} on the disordered phase of the singularity-free $S^2$ nonlinear $\si$-model.  Understanding the properties of its disordered phase represents a long-standing challenge. It is most commonly studied by taking a large $N$ limit, working instead with the $S^N$ nonlinear $\si$-model~\cite{GJ7810} or the  $\mathbb{C}P^{N-1}$ nonlinear sigma model~\cite{DDL7863,W7985}. Interestingly, these two large $N$ limits yield different results for the disordered phase of the nonlinear $\si$-model.  

The disordered state of large $N$ $S^N$ nonlinear $\si$-model is found to be a
gapped symmetric state with no degenerate ground state. This is consistent with
the result of this paper. Since ${\pi_n(S^N) = 0}$ for ${n=0,1\cdots,N-1}$, the
$S^N$ nonlinear $\si$-model has no generalized symmetry in singularity-free
limit for large $N$.  Thus, its disordered state is the gapped product state.
On the other hand, the disordered state of the $\mathbb{C}P^{N-1}$ nonlinear
$\si$-model is a gapless state described by $U(1)$ gauge theory. This is also
consistent with the result of this paper. Since ${\pi_2(\mathbb{C}P^{N-1}) =
\Z}$ and ${\pi_n(\mathbb{C}P^{N-1}) = 0}$ for ${n=0,1,3,4,5, \cdots, 2N-2}$,
the large $N$ $\C P^{N-1}$ nonlinear $\si$-model in singularity-free limit has
a $U(1)$ 1-form symmetry.  Its disordered state spontaneously breaks this
$U(1)$ 1-form symmetry, which produces a state with gapless mode described by a
$U(1)$ gauge field~\cite{GW14125148}.  

The two distinct results obtained from these different large-$N$ approaches
were, by some, interpreted as representing two distinct possible disordered
phases of the singularity-free $S^2$ nonlinear $\si$ model. However, in light
of the findings presented in this paper, this interpretation is incorrect.

Using the formalism we develop in this paper, we non-perturbatively determine
the disordered phase of the singularity-free $S^2$ nonlinear $\si$ model. In
${2+1}$ dimensional spacetime, because ${\pi_0(S^2) = \pi_1(S^2) = 0}$ while
${\pi_2(S^2) = \Z}$, we find that the disordered phase in the singularity free
limit is the Coulomb phase of a $U(1)$ gauge theory. This disagrees with the
large $N$ analysis of the $S^N$ model but agrees with that of the $\C P^{N-1}$
model. However, in $3+1$ dimensions, we find that both of these large $N$
limits of $S^N$ and  $\C P^{N-1}$ fail to capture the correct disordered phase.
These large $N$ limits fail because in $3+1$ dimensions, the singularities
arise from ${\pi_2(S^2) = \Z}$ \textit{and} ${\pi_3(S^2) = \Z}$. In particular,
the singularity-free $S^2$ nonlinear $\sigma$-model in ${(3+1)}$D has a
$3\text{-}\cRep(\mathbb{G}^{(3)})$ symmetry, where the 3-group
$\mathbb{G}^{(3)}$ captures the homotopy 3-type of $S^2$ (the data ${\pi_2(S^2)
= \pi_3(S^2)  = \Z}$ as well as the non-trivial interplay between $\pi_2(S^2)$
and $\pi_3(S^2)$, i.e., a nontrivial Postnikov invariant). Therefore, the
$3\text{-}\cRep(\mathbb{G}^{(3)})$ symmetry is an exotic non-invertible
symmetry, and the above two large-$N$ results do not apply to $S^2$ nonlinear
$\si$-model since they give rise to wrong emergent symmetries. We find that due
to this non-invertible symmetry, the disordered phase of singularity-free $S^2$
nonlinear $\sigma$-model in 3+1D is described by axion electrodynamics,
featuring two gapless modes: one corresponding to a photon and the other to a
massless axion.

\section{Discrete nonlinear ${\si}$-models}\label{dismodel}

Quantum nonlinear ${\si}$-models are widely used in quantum field theory to
describe bosonic quantum systems and condensed matter theory to describe the
dynamics of order parameters in a spontaneous symmetry-breaking state. However,
a nonlinear ${\si}$-model as a quantum field theory is not well defined, at
least when we add nonlinear perturbations, since the path integral that defines
the nonlinear ${\si}$-model requires a summation over $\infty^\infty$
configurations of continuous fields. To obtain a well-defined theory, we
discretize both the ${(d+1)}$-dimensional spacetime $M^{d+1}$ and the target
space $K$. We replace them with simplicial complexes $\cM^{d+1}$ and $\cK$. The
path integral becomes a finite sum in this case, and the discrete quantum
nonlinear ${\si}$-model is well defined. Furthermore, using a lattice
regularization allows us to study the nonlinear ${\si}$-model at strong
coupling, where the physics will likely no longer describe the dynamics of
order parameters in spontaneous symmetry-breaking phases.

\subsection{Simplicial complexes and simplicial sets}\label{SimpCompSec}

\begin{figure}[t!]
\centering
    \includegraphics[width=.48\textwidth]{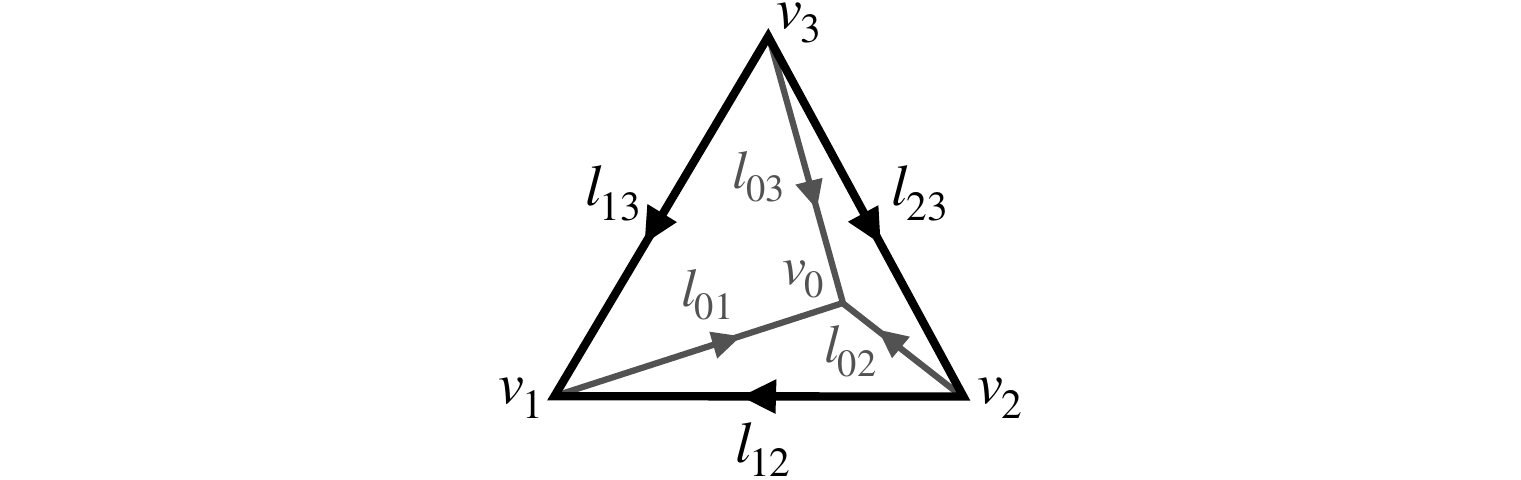}
    \caption{A tetrahedron with a branch structure. A branching structure is a choice of orientation for each link in the complex so that there is no oriented loop on any triangle.} \label{tetr} 
\end{figure}

To describe the discrete nonlinear ${\si}$-model in detail, let us first
briefly review simplicial complexes. A simplicial complex is a set of vertices,
links, triangles, \etc, along with a required set of relations between
cells of different dimensions. We will denote ${K_0,K_1,K_2,\cdots, K_{d+1}}$
as the sets of vertices, links, triangles, \etc\ up to ${d+1}$ cells,  that
form the target space complex $\cK$, and ${M_0,M_1,M_2,\cdots, M_{d+1}}$ as the
sets of vertices, links, triangles, \etc\ that form the spacetime complex
$\cM^{d+1}$. Furthermore, We will use ${v_1,v_2,\cdots \in K_0}$ to label
different vertices in the complex $\cK$, ${l_1,l_2,\cdots \in K_1}$ to label
different links in the complex $\cK$, and ${t_1,t_2,\cdots \in K_2}$ different
triangles, \etc. In addition to the ${n}$-simplicies making up each of these
simplicial complexes, they also have a branching structure (e.g., Fig.
~\ref{tetr}). For the spacetime complex $\cM^{d+1}$, we use this branching
structure to label their cells. We will use $i$ to label vertices in $M_0$,
${(ij)}$ to label links in $M_1$ that connect vertices ${i}$ and ${j}$, and
${(ijk)}$ to label triangles in $M_2$, \etc.

As mentioned, the vertices, links, triangles, \etc\ are related. Those
relations are formally described by
\begin{equation}\label{eq:nerveM}
\xymatrix{ 
M_0 & 
M_1 \ar@<-1ex>[l]_{d_0, d_1}\ar[l] & 
M_2 \ar@<-1ex>_{d_0, d_1 , d_2}[l] \ar@<1ex>[l] \ar[l] & 
M_3 \ar@<-1ex>[l]_{d_0, ..., d_3} \ar@<1ex>[l]_{\cdot} & 
M_4 \ar@<-1ex>[l]_{d_0, ..., d_4} \ar@<1ex>[l]_{\cdot} \cdots ,
}
\end{equation}
where $d_i$ are the face maps, describing how the ${(n-1)}$-simplices are attached to a $n$-simplex. For example
\begin{align}
 d_0(ij) = j,\ \ \
 d_1(ij) = i,
\end{align}
indicates that the tail of the oriented link ${(ij)}$ is connected to the vertex $j$, and the head of the link is connected to the vertex $i$. Similarly, the target space complex $\cK$ is formally described by
\begin{equation}\label{eq:nerveT}
\xymatrix{ 
K_0 & 
K_1 \ar@<-1ex>[l]_{d_0, d_1}\ar[l] & 
K_2 \ar@<-1ex>_{d_0, d_1 , d_2}[l] \ar@<1ex>[l] \ar[l] & 
K_3 \ar@<-1ex>[l]_{d_0, ..., d_3} \ar@<1ex>[l]_{\cdot} & 
K_4 \ar@<-1ex>[l]_{d_0, ..., d_4} \ar@<1ex>[l]_{\cdot} \cdots.
}
\end{equation}  
For example
\begin{align}
 d_0(l_{12}) = v_2,\ \ \
 d_1(l_{12}) = v_1,
\end{align}
indicates that the tail of the oriented link $l_{12}$ is connected to the vertex $v_2$, and the head of the link is connected to the vertex $v_1$ (see Fig.~\ref{tetr}).

A more abstract notion than a simplicial complex is what is known as a simplicial set. Like a simplicial complex, a simplicial set is an object made up of simplicies (vertices, links, triangles, \etc), but its definition is less restrictive than that of a simplicial complex (see Appendix~\ref{SSet}). While ${\cK}$ is a triangulation of ${K}$, we can also consider a simplicial set ${\cB}$ of which ${K}$ is the realization. This means that given ${\cB}$, there is a procedure for producing the topological space $K$ by using the data of $\cB$ to glue together simplices. We will often call $\cB$ a \emph{simplicial-set triangulation} of our target space $K$, but a warning is necessary here since the realization of $\cB$ may not be triangulated in the technical sense and the realization of ${K}$ from ${\cB}$ is only a CW-complex. But, because ${K = |\cB|}$ is built by gluing together vertices, links, triangles, etc., we still use the word ``triangulation'' and the qualifier ``simplicial-set'' is meant to remind the reader that this is not strictly a triangulation in the mathematical sense of the term.

\subsection{Discrete nonlinear ${\si}$-model}\label{sec:discreteNLSM}

A \emph{discrete nonlinear ${\si}$-model} is defined via the following path integral 
\begin{align} \label{discNLSM}
Z(\cM^{d+1};\cK,\cL)=\sum_{\vphi} \ee^{- \int_{\cM^{d+1}} \cL(\vphi)},
\end{align} 
where $\sum_{\vphi}$ sums over all the maps ${\vphi\colon \cM^{d+1}\to \cK}$, and
\begin{align}
 \ee^{-  \int_{\cM^{d+1}} \cL(\vphi)} = 
\hskip -1em
\prod_{\text{cells of } \cM^{d+1} } 
\hskip -1em
\ee^{-  \int_\text{cell} \cL(\vphi)} .
\end{align}
Throughout this paper, we will always assume that ${\cK}$ is connected. The map $\vphi$ is defined as assigning a label ${v_i\in K_0}$ to each vertex ${i\in M_0}$, a label ${l_{ij}\in K_1}$ to each link ${(ij) \in M_1}$, a label ${t_{ijk}\in K_2}$ to each triangle ${(ijk) \in M_2}$, \etc.  Thus, we can view the map $\vphi$ as a collection of fields on the spacetime complex $\cM$: a field $v_i$ on the vertices $M_0$, a field $l_{ij}$ on the links $M_1$, a field $t_{ijk}$ on the triangles $M_2$, \etc. These are all independent fields with their own independent fluctuations.  Therefore, we can rewrite the path integral~\eqref{discNLSM} as a sum over each of these independent fields:
\begin{equation}
Z(\cM^{d+1};\cK,\cL) =
\hskip -1em
\sum_{v_i,l_{ij},t_{ijk},\cdots} 
\hskip -1em
\ee^{- \int_{\cM^{d+1}} \cL(v_i,l_{ij},t_{ijk},\cdots)  } ,
\end{equation}
where
\begin{equation}
 \ee^{-  \int_{\cM^{d+1}} \cL(v_i,l_{ij},t_{ijk},\cdots)} =  \hskip -1.5em \prod_{\text{cells of } \cM^{d+1} }  \hskip -1.5em  \ee^{-  \int_\text{cell} \cL(v_i,l_{ij},t_{ijk},\cdots)}
\end{equation}
Here, ${\ee^{-  \int_\text{cell} \cL(v_i,l_{ij},t_{ijk},\cdots)}}$ is a local action amplitude because it is defined for each cell ${(i,j,k,\cdots)}$ in the spacetime complex $\cM^{d+1}$ and ${v_i,l_{ij},t_{ijk},\cdots}$ in ${\cL(v_i,l_{ij},t_{ijk},\cdots)}$ are the values of the fields on this cell.

\subsection{Singularity-free discrete nonlinear ${\si}$-model} \label{singularity-freeNLSM}

With the above careful definition of a discrete nonlinear ${\si}$-model, we can now define a singularity-free discrete nonlinear ${\si}$-model, whose name we will explain in a moment. A \emph{singularity-free discrete nonlinear ${\si}$-model} is defined by the path integral 
\begin{align} 
\label{ZMKL} 
Z(\cM^{d+1};\cK,\cL)=\sum_{\phi} \ee^{- \int_{\cM^{d+1}} \cL(\phi)},
\end{align} 
where $\sum_{\phi}$ sums over all the homomorphisms between complexes ${\phi\colon\cM^{d+1}\to \cK}$. This differs from Eq.~\eqref{discNLSM}, which is a sum of all maps ${\varphi}$ instead of simplicial homomorphisms ${\phi}$. A homomorphism between complexes is a special kind of map between complexes that preserves the relation between vertices, links, triangles, \etc\ described by Eqs.~\eqref{eq:nerveM} and~\eqref{eq:nerveT}. For example, if ${i,j}$ are two vertices of ${\cM^{d+1}}$ that are attached to the link $(ij)$, the homomorphism $\phi$ maps $i,j$ to ${\phi(i),\phi(j)}$, and $(ij)$ to ${\phi( (i,j))}$ in $\cK$ such that ${\phi(i),\phi(j)}$ are attached to the link $\phi( (i,j))$.  In contrast, for a general map $\vphi$, the vertices ${\vphi(i),\vphi(j)}$ may not be attached to the link $\vphi( (i,j))$. Therefore, all simplicial homomorphisms ${\phi}$ are maps ${\varphi}$, but not vice versa.

The simplicial homomorphism condition is equivalent to the smoothness condition
of a continuum field that maps between spaces ${\phi(x)\colon M^{d+1}\to K}$.
Therefore, the singularity-free discrete nonlinear ${\si}$-model is a lattice
regularization of the continuum nonlinear ${\si}$-models studied in field
theory without singularities. Whereas the continuum field theory is not well
defined, the lattice regularized path integral is.

Physically, the homomorphism (smoothness) condition describes configurations without any dynamical singularities (see footnote~\ref{Singularityfootnote}). Indeed, maps ${\vphi}$ with singularities are those that fail to preserve the attachment structure of the complexes as described by the face maps~\eqref{eq:nerveM} and~\eqref{eq:nerveT}~\cite{ZW180809394}. Therefore, while singularities are present in the discrete nonlinear ${\si}$-model~\eqref{discNLSM}, they are absent in model~\eqref{ZMKL}, which is why we call it the \textit{singularity-free} discrete nonlinear ${\si}$-model. Nondynamical probe singularities---singularity defects---can be added by hand by modifying the sum in~\eqref{ZMKL} to include some maps violating the homomorphism condition.

Let us remark on the classification of singularities, as it will be important later. It is well known that singularities (and singularity defects) are classified by the homotopy groups ${\pi_n(K)}$ of the target space ${K}$~\cite{M1979591}, with the ${n}$th homotopy group detecting ${(d-n)}$-dimensional (in spacetime) singularities. Therefore, in ${(d+1)}$-dimensional spacetime with connected ${K}$, singularities are classified by $\pi_n(K)$ for ${1\leq n \leq d}$. The obstructions related to $\pi_n(K)$ for ${n\geq d+1}$ do not have a physical interpretation of excitations/defects. So, the classification of singularities in a ${(d+1)}$-dimensional nonlinear ${\si}$-model with target space ${K}$ is the same as if the target space was ${K_{\tau\leq d}}$, which satisfies
\begin{equation}\label{dPosSta}
\pi_n(K_{\tau\leq d}) = \begin{cases}
\pi_n(K)\quad\quad& \text{if }n\leq d, \\
0\quad\quad &\text{if else}.
\end{cases} 
\end{equation}
When ${\pi_n(K)}$ are finite for all ${n\leq d}$, we refer to the nonlinear ${\si}$-model as a $\pi$-finite nonlinear ${\si}$-model, and otherwise we refer to it as a $\pi$-infinite nonlinear ${\si}$-model.

The space ${K_{\tau\leq d}}$ is called a \emph{Postnikov stage} of $K$ (see Appendix~\ref{Postnikov}), and will play an important role in Sec.~\ref{genSymHigherGaugeThy}. In particular, ${K_{\tau\leq d}}$ is the classifying space of a ${d}$-group ${\mathbb{G}^{(d)}}$:
\begin{align}\label{dGrouphomotopyGroup}
K_{\tau\leq d} = B\mathbb{G}^{(d)} =\cB_*\big( \pi_1(K), \pi_2(K),\cdots, \pi_d(K)\big),
\end{align} 
where the subscript $*$ is shorthand for the data of the ${d}$-group that describes the relations between ${\pi_1(K),\cdots,\pi_d(K)}$. As reviewed in Appendix~\ref{hgroup}, this data includes how ${\pi_1(K)}$ acts on ${\pi_n(K)}$, for each ${n \in \{2,\cdots, d\}}$, along with a set of cocycles that describe how singularities of different dimensions are related. For ${\pi}$-finite nonlinear ${\si}$-models, ${\mathbb{G}^{(d)} }$ is a finite ${d}$-group. Otherwise, it is a nonfinite discrete ${d}$-group. The relationship between ${K_{\tau\leq d}}$ and ${\mathbb{G}^{(d)}}$ also means that the classification of singularities is the same as the classification of magnetic defects (i.e., gauge fluxes) of ${\mathbb{G}^{(d)}}$ higher gauge theory~\cite{ZW180809394, P230805730}.

\subsection{Singularity-suppressed discrete nonlinear ${\si}$-model} \label{singularity-supNLSM}

Restricting a general map ${\varphi}$ to a simplicial homomorphism ${\phi}$ can be done locally since the relation between vertices, links, triangles, \etc\ are local relations given by the face maps ${d_i}$. Indeed, these relations can be determined locally by examining a single cell in the spacetime complex. Formally, whether these relations defining the homomorphism condition are satisfied can be checked using a function ${\cA(\varphi)}$, which is defined to be zero for fields satisfying the homomorphism condition while unity otherwise:
\begin{equation}
\hspace{-5pt}\cA(\varphi)  \hspace{-1pt}=\hspace{-1pt} \begin{cases}
0 \quad\quad& \text{if}\hspace{8pt}\begin{matrix} d_0(\varphi((i,j))) \hspace{-1pt}= \varphi(j)\\d_1(\varphi((i,j))) \hspace{-1pt}= \varphi(i)\end{matrix},\hspace{5pt}\cdots\hspace{-1pt},\\
1 \quad\quad& \text{if else}.
\end{cases}
\end{equation}

Using ${\cA(\varphi)}$, the local homomorphism restriction can be implemented in the path integral by including a term that penalizes configurations violating it. Using the decomposition of ${\varphi}$ into ${v_i}$, ${l_{ij}}$, ${t_{ijk}}$, \etc\ discussed in Sec.~\ref{sec:discreteNLSM}, we define the \textit{singularity-suppressed} discrete nonlinear ${\si}$-model by the path integral
\begin{equation} \label{ZMKvltU}
Z = \hskip -1em \sum_{v_i,l_{ij},t_{ijk},\cdots}  \hskip -1em \ee^{- \int_{\cM^{d+1}}  \cL(v_i,l_{ij},t_{ijk},\cdots)   +U \cA(v_i,l_{ij},t_{ijk},\cdots)  } .
\end{equation} 
For ${U\to \infty}$, all observables determined from this path integral satisfy the homomorphism restriction exactly, and the theory is the same as the singularity-free nonlinear ${\si}$-model~\eqref{ZMKL} except written as a path integral over local independent fields. For ${U\to 0}$, this model is the same as the discrete nonlinear ${\si}$-model~\eqref{discNLSM}.

The benefit of this model is that, unlike the discrete nonlinear ${\si}$-model~\eqref{discNLSM}, the ${U}$ term allows us to suppress dynamical singularities while considering a model with local independent fields. Indeed, for very large but finite ${U}$, we expect the emergent long-distance behavior to mimic the ${U\to \infty}$ limit. Therefore, although the homomorphism $\phi$ is not a local field, the universal features of the singularity-free discrete nonlinear ${\si}$-model can be described by local bosonic fields and hence a local bosonic model.

\section{Generalized symmetries in singularity-free $\pi$-finite nonlinear ${\si}$-models} 
\label{emsymm}

Generalized symmetries commonly emerge from local constraints~\cite{PW230105261}. Therefore, the local constraint that implements the simplicial homomorphism condition in the singularity-suppressed discrete nonlinear ${\si}$-model may give rise to emergent generalized symmetries. Such generalized symmetries would emerge below the singularity energy scale ${U}$. Crucially, they would be \textit{exact emergent symmetries} when they are higher-form symmetries~\cite{PW230105261} and can, therefore, characterize the model's phases at large finite ${U}$. On the other hand, these generalized symmetries would be exact in the ${U\to \infty}$ limit and, therefore, be exact symmetries of the singularity-free discrete nonlinear ${\si}$-model. 

Generalized symmetries in nonlinear ${\si}$-models have been explored for both particular~\cite{GW14125148, GP180103199, HI180209512, AJ190801175, DM190806977, B201200051, CT221013780, H221208608} and general target spaces ${K}$~\cite{P230805730,CT230700939}. In this section, we will discuss these symmetries of the singularity-free $\pi$-finite nonlinear ${\si}$-model, and in the next section, we'll consider the singularity-free $\pi$-infinite nonlinear ${\si}$-model. Here, we will first find the symmetry category describing it following~\Rf{P230805730}. It is not a rigorous derivation but provides valuable intuition into the mathematical structure describing the symmetry. We then perform a rigorous calculation of the symmetry category, which is one of our main results. To do so, we will use the Symmetry topological order\footnote{SymTO was first called categorical
symmetry~\cite{JW191213492, KZ200514178}, but later renamed since many use categorical symmetry to refer to non-invertible symmetry. SymTO has also been called symmetry topological field theory (SymTFT)~\cite{ABE211202092, A220310063, KOZ220911062, BS230517159}, holographic categorical symmetry~\cite{LZ230512917}, topological holography~\cite{MT220710712}, and topological symmetry~\cite{FT220907471}.} (SymTO)~\cite{KZ150201690, JW190513279,TW191202817,JW191213492, KZ200514178,KZ201102859, KZ220105726,CW220506244}. Therefore, we will first briefly review the SymTO before discussing the symmetry category of the singularity-free $\pi$-finite nonlinear ${\si}$-model. We refer the reader to Sec. 2 of \Rf{CW221214432} for a more comprehensive review.

\subsection{Review of SymTO}

The SymTO is a unified framework for describing \textit{all} generalized symmetries, including their symmetry defects, symmetry charges, and their 't Hooft anomalies. Here, we will review its properties for finite generalized symmetries described by a fusion $d$-category ${\cR\neq \Sigma \cB}$, where $\cB$ is a non-degenerate braided fusion ${(d-1)}$-category. In this case, the symmetry category is described by excitations on a gapped boundary---topological defects on a topological boundary (at low energy)---of a topological order---the SymTO---in one higher dimension. 

Within the Symmetry/Topological-Order (Symm/TO) correspondence, a system with a symmetry restricted to its symmetric sub-Hilbert space can be exactly simulated by a boundary of the corresponding SymTO in one higher dimension. Adding the aforementioned gapped boundary provides a physical realization of the system's symmetry and, at low energies in the SymTO, allows the entirety of its Hilbert space to be simulated. The excitations on the gapped boundary are described by  ${\cR}$, which is a defining characteristic of this gapped boundary. Therefore, the braided fusion ${d}$-category ${\eM}$ describing the excitations of the SymTO is given by the (Drinfeld) center of ${\cR}$~\cite{KZ150201690}:
\begin{equation}
\eM = \eZ(\cR).
\end{equation}

An important concept related to the SymTO is the notation of \emph{holo-equivalence}~\cite{KZ200514178}. Two symmetries described by the monoidal ${d}$-categories $\cR_1$ and $\cR_2$, respectively, are holo-equivalent if there exists a one-to-one correspondence between $\cR_1$-symmetric systems and the $\cR_2$-symmetric systems such that 
\begin{enumerate}
\item the corresponding systems' local symmetric operators have identical correlations and
\item the corresponding systems have identical energy spectra when restricted to their respective symmetric sub-Hilbert spaces.
\end{enumerate}
This implies that two symmetries are holo-equivalent if gauging one symmetry yields a new theory with the other symmetry. For finite symmetries, where ${\cR_1}$ and ${\cR_2}$ are both fusion ${d}$-categories, this implies that the two symmetries have the same SymTO~\cite{KZ200514178} and therefore
\begin{align}
 \eZ(\cR_1) = \eZ(\cR_2).
\end{align}
This result for finite symmetries motivated the term \emph{holo-equivalence}. Using mathematical terminology, ${\cR_1}$ and ${\cR_2}$ are holo-equivalent if they are Morita equivalent. The notion of holo-equivalence and SymTO were widely used in \Rfs{HV14116932,CZ190312334,JW191213492,KZ200514178,JW210602069,CW220303596,CW220506244,CW221214432,PW230105261}, where more discussions and applications can be found.

\subsection{The symmetry category}\label{genSymHigherGaugeThy}

The singularity-free nonlinear ${\si}$-model has conserved quantities corresponding to the conservation of ``smooth soliton'' numbers, which come from topologically nontrivial maps from
spacetime to the target space. These conserved quantities, arising from the
absence of singularities, label the topological sectors of the
singularity-free discrete nonlinear ${\si}$-model (and the continuum nonlinear
${\si}$-model).  From the point of view of generalized symmetries, these
topological sectors are symmetry sectors. So, what is this generalized
symmetry?

Since singularities take configurations from one topological sector to another,
they are the charged ``operators'' of the generalized
symmetry~\cite{P230805730}. Recall from Sec.~\ref{singularity-freeNLSM} that
the classification of singularities is the same as the classification of
magnetic defects of ${\mathbb{G}^{(d)}}$ higher gauge theory (see
Eq.~\eqref{dGrouphomotopyGroup}). Therefore, this symmetry is the same as the
magnetic symmetry of ${\mathbb{G}^{(d)}}$ higher gauge theory. The symmetry
defects obey the same description as the ${\mathbb{G}^{(d)}}$ electric defects
(e.g., Wilson loops) and, therefore, are described by the ${d}$-representations
of ${\mathbb{G}^{(d)}}$. For ${\pi}$-finite ${K}$, this means that
${\mathbb{G}^{(d)}}$ is finite and the symmetry category is
\begin{equation}\label{symCat1}
\cR = d\text{-}\cRep(\mathbb{G}^{(d)}),
\end{equation}
the fusion ${d}$-category of ${d}$-representations of ${\mathbb{G}^{(d)}}$.  By
definition, this fusion $d$-category is the $d$-functor category
\begin{equation}
\cR = [K_{\tau\leq d}, d\text{-}\cVec].
\end{equation}
Because this symmetry depends only on the topology of the field's target space,
it cannot be broken or modified by local modifications to the action.

We see that the description of this symmetry is quite complicated and that it is generally a non-invertible higher-form symmetry. However, using the notion of dual symmetry introduced in \Rf{JW191213492}, we can have a simpler description: the non-invertible ${d\text{-}\cRep(\mathbb{G}^{(d)})}$ symmetry is the dual of a higher-group symmetry described by the $d$-group $\mathbb{G}^{(d)}$, or equivalently by the fusion ${d}$-category
\begin{equation}
\tl\cR = d\text{-}\cVec_{\mathbb{G}^{(d)}}.
\end{equation}
${\cR}$ and ${\tl\cR}$ being dual symmetries implies they are holo-equivalent. The anyons of their SymTOs are described by the braided fusion ${d}$-category ${ \eZ(\cR) =  \eZ(\tl\cR)}$, making their SymTOs $\mathbb{G}^{(d)}$ higher gauge theory in ${d+2}$ dimensional spacetime (\ie in one higher dimension). Therefore, phases and phase transitions of the singularity-free ${\pi}$-finite nonlinear ${\si}$-model characterized by ${\cR}$ can be mapped onto phases and transitions of a dual model characterized by ${\tl\cR}$.

When ${d=2}$, this generalized symmetry is always holo-equivalent to a
(potentially anomalous) finite ordinary symmetry. Indeed, one can first gauge
${\cR}$ to get the dual symmetry ${\mathbb{G}^{(2)}}$ and then gauge the 1-form
symmetry of ${\mathbb{G}^{(2)}}$ to get an ordinary finite symmetry ${G}$. This
means that the SymTO of ${\cR}$ for ${d=2}$ will always be ${(3+1)}$D ${G}$
gauge theory. This is consistent with the conjecture that ${3+1}$D bosonic
topological orders without emergent fermions are described by Dijkgraaf-Witten
theories of finite ordinary groups~\cite{LW170404221}.

Let us make two additional remarks about the generalized symmetry ${\cR}$. 

Firstly, the SymTO is a finite higher gauge theory without any cocycle twist
${[\om]\in H^{d+2}(B\mathbb{G}^{(d)}, U(1))}$. However, if our singularity-free
${\pi}$-finite discrete nonlinear ${\si}$-model had a Wess-Zumino-Witten (WZW)
term~\cite{WZ7195, W8322}, then the SymTO will be a higher Dijkgraaf-Witten
theory~\cite{thorngrenThesis}. Indeed, a WZW term would have a local expression
in one higher dimension, appearing as the action amplitude ${[\om]\in
H^{d+2}(K_{\tau\leq d}, U(1))}$ pulled back to ${(d+2)}$-dimensional spacetime.
This would modify the SymTO action amplitude by this pulled-back cocycle,
causing it to describe twisted ${\mathbb{G}^{(d)}}$ higher gauge theory--a
higher Dijkgraaf-Witten theory---because of Eq.~\eqref{dGrouphomotopyGroup}.
This means that Eq.~\eqref{symCat1} will not describe the generalized
symmetries of nonlinear ${\si}$-models with a WZW term. What the symmetry
category ${\cR}$ becomes with a WZW term is nontrivial. However, in terms of
the dual symmetry ${\tl\cR}$, it is much simpler: the emergent symmetry will be
the dual of an anomalous $\mathbb{G}^{(d)}$ symmetry described by ${\tl\cR =
d\text{-}\cVec^\om_{\mathbb{G}^{(d)}}}$.

Secondly, the target space $K$ of a continuous $\pi$-finite nonlinear
$\si$-model is usually a symmetric space with a symmetry described by a group
$G_\text{symm}$. Thus far, we have ignored this $G_\text{symm}$ symmetry. If we
included it, then we conjecture that the emergent symmetry (up to
holo-equivalence) of the singularity-free continuous $\pi$-finite nonlinear
${\si}$-model is described by a ${d}$-group $\cPSG^{(d)}$ defined by the
nontrivial extension
\begin{align}\label{PSGext}
0 \to \mathbb{G}^{(d)} \to \cPSG^{(d)} \to G_\text{symm} \to 0 .
\end{align}
This conjecture is motivated by the projective symmetry group (PSG) result from~\Rf{W0213}, where $\mathbb{G}^{(d)}$ and ${\cPSG^{(d)}}$ are both ordinary groups. By gauging ${\mathbb{G}^{(d)}}$ in ${\cPSG^{(d)}}$, this is holo-equivalent to the conjecture made in~\Rf{P230805730} that the emergent symmetry ${\cR}$ and the microscopic symmetry ${G_{\text{symm}}}$ have a mixed 't Hooft anomaly. The data specifying the nontrivial extension of ${G_{\text{symm}}}$ by ${\mathbb{G}^{(d)}}$ in Eq.~\eqref{PSGext} is the same data specifying the nontrivial mixed anomaly between ${G_{\text{symm}}}$ and ${\cR}$.

\subsection{A rigorous calculation based on SymTO}\label{rigorCalcSymTO}

The above discussion on the symmetry category ${\cR = d\text{-}\cRep(\mathbb{G}^{(d)})}$ and its SymTO being ${\mathbb{G}^{(d)}}$ higher gauge theory was based on the homotopy ${n}$-type of the target space ${K}$ and not an explicit calculation using the discrete nonlinear ${\si}$-model's path integral. In this section, we give a more explicit and rigorous derivation of this result, showing that the ${(d+1)}$-dimensional singularity-free ${\pi}$-finite discrete nonlinear ${\si}$-model can be viewed as a boundary of ${(d+2)}$-dimensional $\mathbb{G}^{(d)}$ higher gauge theory. This result implies that the SymTO of ${\cR}$ is indeed $\mathbb{G}^{(d)}$ higher gauge theory.

The partition function of the singularity-free nonlinear ${\si}$-model is given by Eq.~\eqref{ZMKL}. We will assume that spacetime $\cM^{d+1}$ is a ${(d+1)}$-sphere: ${\cM^{d+1} = \cS^{d+1}}$. Because we are interested only in the singularity defects of the model, without a loss of generality, we truncate the connected target space ${K}$ to the ${d}$th Postnikov stage ${K_{\tau\leq d}}$, which satisfies Eq.~\eqref{dPosSta}. We will denote the realization of ${K}$ from the simplicial set ${K_{\tau\leq d}}$---the simplicial set triangulation of ${K}$---as ${\cB}$.

To show that the singularity-free nonlinear ${\si}$-model with target space ${\cB}$ can be viewed as a boundary of $\mathbb{G}^{(d)}$-gauge theory, it is enough to show that the action amplitude in the path integral \eqref{ZMKL} on $\cS^{d+1}$ can be rewritten as
\begin{equation} \label{actSKL} 
\ee^{- \int_{\cS^{d+1}} \cL(\phi)}= \ee^{- \int_{\cS^{d+1}} \cL(\phi)}C \sum_{\phi^D} 1,
\end{equation}
where ${\sum_{\phi^D} 1}$ is a path integral on a ${(d+2)}$-dimensional disk $\cD^{d+2}$ that sums over all simplicial homomorphisms ${\phi^D\colon \cD^{d+2} \to \cB}$ with the fixed boundary conditions
\begin{equation}\label{SymTOBC}
\phi^D\bigg|_{\prt \cD^{d+2} = \cS^{d+1}} = \phi\colon \cS^{d+1} \to \cB,
\end{equation}
and $C$ is a $\phi$ independent constant. Since ${\cB}$ models the classifying space of the ${d}$-group ${\mathbb{G}^{(d)}}$ (see Eq.~\eqref{dGrouphomotopyGroup} and appendices~\ref{hgroup} and~\ref{Postnikov}), ${\sum_{\phi^D} 1 }$ describes the deconfined phase of $\mathbb{G}^{(d)}$-gauge theory in ${(d+2)}$-dimensional spacetime ${\cD^{d+2}}$. Therefore, Eq.~\eqref{actSKL} shows that the singularity-free nonlinear ${\si}$-model can be viewed as a boundary of $\mathbb{G}^{(d)}$-gauge theory. Because topological defects on the Neumann boundary of $\mathbb{G}^{(d)}$-gauge theory are described by ${d\text{-}\cRep(\mathbb{G}^{(d)})}$, this implies that the $\mathbb{G}^{(d)}$-gauge theory can be used to describe the SymTO for the symmetry~\eqref{symCat1}.

To prove \eqref{actSKL}, we note that ${\sum_{\phi^D} 1}$ counts the number of simplicial homomorphisms ${\phi^D\colon \cD^{d+2} \to \cB}$ that obey the boundary condition~\eqref{SymTOBC}. Eq.~\eqref{actSKL} means that the number of these simplicial homomorphisms is independent of the boundary condition.  

Since ${\cB}$ is the classifying space of $\mathbb{G}^{(d)}$, the sum is over all flat $\mathbb{G}^{(d)}$ gauge fields satisfying the boundary conditions. These gauge fields are a collection of cochains ${a^{1}\in Z^1(\cD^{d+2} ,\pi_1(K))}$ and ${a^{i}\in C^i(\cD^{d+2} ,\pi_i(K)^{\al_i})}$ (${2\leq i\leq d}$) satisfying the flatness conditions (see appendix~\ref{hgroup})
\begin{equation}\label{higherGrpCocycleCond}
\begin{aligned}
a^{1}_{ij} a^{1}_{jk} &= a^{1}_{ik}\\
\dd_{\al_{i}} a^{i} &= c_{i+1}(a^{1},\cdots,a^{i-1}),
\end{aligned}
\end{equation}
where ${c_{i+1}}$ is a ${\pi_i(K)}$ ${(i+1)}$-cocycle. Since $\cD^{d+2}$ has trivial topology, these flatness conditions can always be solved by considering pure gauge fluctuations.

Let us first assume that only $\pi_1(K)$ is nonzero. In this case, the ${d}$-group ${\mathbb{G}^{(d)}}$ is equivalent to the ${1}$-group $\pi_1(K)$, so ${\sum_{\phi^D} 1 }$ describes $\pi_1(K)$ gauge theory and is a sum over all flat $\pi_1(K)$-connections satisfying~\eqref{SymTOBC}. In particular, ${\phi^D}$ consists of only 1-cochains ${a^{1}_{ij}\in \pi_1(K)}$ on the links ${(ij)}$ in ${\cD^{d+2}}$ satisfying the flatness condition ${a^{1}_{ij}a^{1}_{jk} = a^{1}_{ik}}$ with ${a^{1}_{ij}}$ fixed for links on ${\prt \cD^{d+2}}$. Therefore, in this case
\begin{equation}
\begin{aligned}
& \sum_{\phi^D} 1 = \sum_{a^{1}_{ij},\ (ij) \in \cD^{d+2}_\text{int} } 1,\\
& \text{ where } \ a^{1}_{ij} a^{1}_{jk}=a^{1}_{ik}, \ \ \ \ a^{1}_{ij} \in \pi_1(K),
\end{aligned}
\end{equation}
where ${\cD^{d+2}_\text{int}}$ is the interior of ${\cD^{d+2}}$---is
${\cD^{d+2}}$ minus simplicies belonging to ${\prt\cD^{d+2}}$. On the disk, the
cocycle condition ${a^{1}_{ij} a^{1}_{jk}=a^{1}_{ik}}$ is solved by
${a^{1}_{ij}=f^{(0)}_i(f^{(0)}_j)^{-1}}$, where ${f^{(0)}_i \in \pi_1(K)}$.
We notice that while ${(ij)}$ in ${a^{1}_{ij}=f^{(0)}_i(f^{(0)}_j)^{-1}}$ are
links of ${\cD^{d+2}_\text{int}}$, ${i}$ and ${j}$ are vertices of
${\cD^{d+2}}$ (including ${\prt\cD^{d+2}}$).  Since we fix ${a^{1}_{ij}}$
on the boundary ${\cD^{d+2}_\text{int}}$, this requires us to fix ${f^{(0)}_i}$
on the boundary ${\cD^{d+2}_\text{int}}$. Using this solution, the sum
evaluates to
\begin{align}
& \sum_{\phi^D} 1 = |\pi_1(K)|^{-1}\hspace{-21pt}\sum_{\hspace{15pt}f^{(0)}_i,~i\in 
\cD^{d+2}_\text{int}} \hspace{-7pt}1 \hspace{3pt} =  |\pi_1(K)|^{N_0-1},
\end{align}
where ${N_j}$ is the number of ${j}$-simplicies in $\cD^{d+2}_\text{int}$. The
factor ${ |\pi_1(K)|^{-1}}$ is included to remove overcounting coming from
${f^{(0)}_i}$ and ${\tl f^{(0)}_i = f^{(0)}_i g}$ giving rise to the same
${a^{1}_{ij}}$ for any ${g \in \pi_1(K)}$. We see that ${\sum_{\phi^D} 1}$ is
independent of the boundary condition~\eqref{SymTOBC}, thus proving
Eq.~\eqref{actSKL} for this case.

Next, we assume that only $\pi_1(K)$ and $\pi_2(K)$ are non-zero. In this case,
${\mathbb{G}^{(d)}}$ is equivalent to a 2-group and the sum
\begin{align}\label{pi1pi2SimCount}
& \sum_{\phi^D} 1 =
 \sum_{a^{1}_{ij}, a^{2}_{ijk}\ (ij), (ijk) \in \cD^{d+2}_\text{int} } 1 
\\
& \text{ where } \ 
a^{1}_{ij} a^{1}_{jk}=a^{1}_{ik}, \ \ \ \ a^{1}_{ij} \in \pi_1(K).
\nonumber\\
&\ \ \ \ \ \ \ \ \
(\dd_{\al_2} a^{2})_{ijkl}
=
c_3(a^{1})_{ijkl} , \ \ \ \ a^{2}_{ijk} \in \pi_2(K).
\nonumber 
\end{align}
Again, the sum is only over cochains on links $(ij)$ and triangles $(ijk)$ in
the interior of ${\cD^{d+2}}$ since the boundary condition fixes ${a^{1}}$
and ${a^{2}}$ on ${\prt\cD^{d+2}}$. The pure gauge fluctuations solving the
flatness condition is
\begin{align}
a^{1}_{ij} = (f^{(0)})_i (f^{(0)})_j^{-1},\quad
a^{2} = \dd_{\al_2} f^{(1)} + \nu^{(2)}(f^{(0)}),
\end{align}
where 
${f^{(0)} \in \pi_1(K)}$, 
${f^{(1)} \in \pi_2(K)}$, 
and ${\nu^{(2)}}$ is a first descendant of ${c_3}$~\cite{thorngrenThesis},
\ie  ${\nu^{(2)}}$ is a solution of 
\begin{align}
{\dd_{\al_2} \nu^{(2)}(f^{(0)}) = c_3(\dd f^{(0)})} 
\end{align}
on $ \cD^{d+2}$. We want to use ${(f^{(0)},f^{(1)})}$ to
label ${(a^{1},a_2^{(2)})}$, but the labeling is many-to-one. Notice that
${f^{(1)}}$ and ${f^{(1)}+ \dd_{\al_2} g^{(0)}}$, with ${g^{(0)}_i\in
\pi_2(K)}$, correspond to the same ${a^{2}}$, but ${g^{(0)}_i}$ and
${g^{(0)}_i+g}$, with ${g\in \pi_2(K)}$ produce the same ${\dd_{\al_2} g^{(0)}}$. This leads to
an overcounting factor of ${|\pi_2(K)|^{N_0-1}}$. We also notice that
${(f^{(0)}_i, f^{(1)})}$ and ${(f^{(0)}_i h, f^{(1)}-\nu^{(1)}(f^{(0)},h))}$, where ${h\in\pi_1(K)}$ and ${\nu^{(1)}(f^{(0)},h)}$ satisfies
\begin{align}
 \dd_{\al_2} \nu^{(1)}(f^{(0)},h) =
 \nu^{(2)}(f^{(0)}h)-
 \nu^{(2)}(f^{(0)}),
\end{align}
give
rise to the same $(a^{1},a^{2})$. The above equation always has a solution on ${\cD^{d+2}} $, since
\begin{equation}
\begin{aligned}
& \dd_{\al_2}  [\nu^{(2)}(f^{(0)}h)-
 \nu^{(2)}(f^{(0)}) ] \\
&\hspace{30pt}= c_3\big(\dd (f^{(0)}h)\big) - c_3\big(\dd f^{(0)}\big) = 0. 
\end{aligned}
\end{equation}
This leads to an overcounting factor of $|\pi_1(K)|$. Thus, we can
replace the sum over $a^{1}_{ij}$ and $a^{2}_{ijk}$
in~\eqref{pi1pi2SimCount} with a sum over $(f_1)_{i}$ and $(f_2)_{ij}$ and
obtain
\begin{align}\label{pi1pi2SimCount2}
& \sum_{\phi^D} 1 =
|\pi_2(K)|^{-N_0+1} 
|\pi_1(K)|^{-1} 
\sum_{\substack{f^{(0)}_{i},~i \in \cD^{d+2}_\text{int}\\f^{(1)}_{ij},~ (ij) \in \cD^{d+2}_\text{int}}} 1.
\end{align}
Computing this sum, we find
\begin{align}
& \sum_{\phi^D} 1 = 
|\pi_2(K)|^{N_1-N_0+1}|\pi_1(K)|^{N_0-1},
\end{align}
and again see that $\sum_{\phi^D} 1$ is independent of the boundary condition of $\phi^D$.

The same calculation can be done for general case. It is tedious but straightforward and shows that the partition function of the singularity-free nonlinear
${\si}$-model~\eqref{ZMKL} on $\cS^{d+1}$ can be rewritten as a path integral
on a $d+2$-dimensional disk $\cD^{d+2}$:
\begin{align}
Z(\cS^{d+1};\cK,\cL) &=
\sum_{\phi} \ee^{- \int_{\cS^{d+1}} \cL(\phi)} 
\nonumber\\
&
= 
\sum_{\phi^D} C \ee^{- \int_{\cS^{d+1}} \cL(\phi^D|_{\cS^{d+1}})} .
\end{align}
The path integral $\sum_{\phi^D}$ on $\cD^{d+2}$ describes the deconfined phase
of $\mathbb{G}^{(d)}$ higher-gauge theory.  Thus, the singularity-free
nonlinear ${\si}$-model can be viewed as a boundary of the deconfined phase of
$\mathbb{G}^{(d)}$ higher-gauge theory. This implies that the SymTO for its
generalized symmetries is $\mathbb{G}^{(d)}$ higher-gauge theory. We emphasize
how this is true for arbitrary choices of $ \cL(\phi)$, as long as the
singularity-free condition is satisfied.

\section{Generalized symmetries in singularity-free $\pi$-infinite nonlinear
${\si}$-models} \label{piinf}

In Sec.~\ref{genSymHigherGaugeThy}, we found that the generalized symmetries of
the singularity-free nonlinear ${\si}$-model in ${(d+1)}$-dimensional spacetime
are described by the ${d}$-representations of the ${d}$-group
${\mathbb{G}^{(d)}}$. We discussed the singularity-free ${\pi}$-finite
nonlinear ${\si}$-model---the situation where ${\mathbb{G}^{(d)}}$ was a finite
${d}$-group---in Sec.~\ref{emsymm}. In this section, we discuss the generalized
symmetry of the singularity-free $\pi$-infinite nonlinear ${\si}$-model. For
physically relevant ${K}$, this means that some $\pi_n(K)$'s (${n\leq d}$)
contain one or more $\Z$ factors, causing ${\mathbb{G}^{(d)}}$ to be a
discrete, but nonfinite ${d}$-group.

Since singularity defects are charged under the generalized symmetry, there are
a countably infinite number of symmetry charges in the singularity-free
$\pi$-infinite nonlinear ${\si}$-model. This causes the generalized symmetry to
be continuous and includes non-invertible parts for
generic target spaces ${K}$. What mathematical structure describes the
corresponding symmetry category is an open question. Furthermore, it is still
being determined what theory takes the role of the SymTO since it seemingly
must describe an infinite braided fusion $d$-category. For the case of ${\Z}$
symmetry charges, a first guess would be ${U(1)}$ gauge theory in one higher
dimension whose ${U(1)}$-connection is always flat. Such gauge theories were
discussed in, for instance, \Rfs{CON12110564, MD230801765}.

While its mathematical description is unknown, we can still discuss some
general aspects of this generalized symmetry. For instance, the generalized
symmetry will be invertible only when there exists a ${d}$-group
${\hat{\mathbb{G}}^{(d)}}$ that is Pontryagin dual to ${\mathbb{G}^{(d)}}$. In
fact, when ${\hat{\mathbb{G}}^{(d)}}$ exists, it will be precisely the
${d}$-group that describes the symmetry defects. Whether or not
${\hat{\mathbb{G}}^{(d)}}$ exists depends on
${\mathbb{G}^{(d)}}$.\footnote{When ${d=1}$, ${\mathbb{G}^{(1)}}$ is an
ordinary group, and ${\hat{\mathbb{G}}^{(1)}}$ exists only when
${\mathbb{G}}^{(1)}$ is abelian.}

Let us first consider the simplest scenario where all group homomorphisms
${\al_n}$ and Postnikov classes ${c_{n+1}}$ (${n\leq d}$) defining
${\mathbb{G}^{(d)}}$ are trivial and ${\pi_1(K)}$ is abelian. In this case,
${\mathbb{G}^{(d)}}$ is a trivial ${d}$-group, given by ``direct products''
${\mathbb{G}^{(d)} =\prod_{r=0}^{d-1}G^{(r)}}$ with ${G^{(r)} = \pi_{r+1}(K)}$.
The dual ${d}$-group ${\hat{\mathbb{G}}^{(d)}}$ is then also a trivial
${d}$-group~\cite{P230805730}, given by ${\hat{\mathbb{G}}^{(d)}
=\prod_{r=0}^{d-1}\hat{G}^{(r)}}$ where ${\hat{G}^{(r)} =
\Hom(\pi_{d-r}(K),U(1))}$ is the Pontryagin dual of ${\pi_{d-r}(K)}$.
Therefore, in this scenario, whenever there is a ${\Z\subset \pi_n(K)}$, there
will be a ${U(1)}$ ${(d-n)}$-form symmetry.

Next, let's consider the case where ${\al_n}$ are trivial but there are
nontrivial Postnikov invariants ${c_{n+1}}$ (${n\leq d}$). The dual higher
group ${\hat{\mathbb{G}}^{(d)}}$ exists only when the Postnikov invariants are
stable cohomology operations.\footnote{We thank Arun Debray and Hao Xu for
related discussion.} For example, Bockstein homomorphisms are stable cohomology
operations, but some cup products are not.\footnote{We refer the reader to~\cite[Definition 12.3.23]{aguilar2002algebraic} for the definition of stable cohomology operation.} Thus, whenever ${\mathbb{G}^{(d)}}$ has a
Postnikov class that is not a stable cohomology operator, the corresponding
singularity-free nonlinear ${\si}$-model will have a noninvertible
symmetry. A simple case where this occurs is for ${K=S^2}$ with ${d>2}$ (see
Sec.~\ref{S2Section}). Therefore, the singularity-free ${S^2}$ nonlinear
${\si}$-model in ${d+1>3}$ has a noninvertible symmetry.

When any of the ${\al_n}$'s are nontrivial (${n\leq d}$), we expect that
${\hat{\mathbb{G}}^{(d)}}$ does not exist, making the generalized symmetry
non-invertible. Therefore, for ${\pi}$-infinite ${K}$ where the ${\pi_1(K)}$
action on ${\pi_n(K)}$ (${n\leq d}$) is nontrivial, the singularity-free
nonlinear ${\si}$-model will have a non-invertible symmetry. A
simple example is ${K = \mathbb{R}P^2}$, where ${\pi_1(K) \simeq
\Z_2}$, ${\pi_2(K) \simeq \Z}$, and the action of ${\pi_1(K)}$ on ${\pi_2(K)}$
flips the sign of ${\pi_2(K)}$~\cite{VM19772256}. So, a singularity-free
${\mathbb{R}P^2}$ nonlinear ${\si}$-model in ${d+1>2}$ will have a
noninvertible symmetry.

\section{Nontrivial disordered phases} \label{examples}

When terms in the Lagrangian ${\cL(\phi)}$ of the singularity-free discrete
nonlinear ${\si}$-model~\eqref{ZMKL} are large, the ground state  wavefunction
is dominated by the constant map ${\phi_0\colon \cM^{d+1}\to \{\text{pt}\} \in
\cK}$ in the path integral. This signals the ordered phase of the nonlinear
${\si}$-model and that the generalized symmetry ${\cR}$ we have been discussing
is not spontaneously broken, causing the singularities to be confined. When the
generalized symmetry is finite, gauging ${\cR=
d\text{-}\cRep(\mathbb{G}^{(d)})}$ maps the nonlinear ${\si}$-model to its dual
model, which will lie in a phase where the entire ${d}$-group
${\mathbb{G}^{(d)}}$ symmetry is spontaneously broken. This symmetry-breaking
pattern is encoded by the electric boundary of the SymTO, whose fusion
${d}$-category is ${\tl\cR= d\text{-}\cVec_{\mathbb{G}^{(d)}}}$.
 
On the other hand, by choosing ${\cL(\phi) = 0}$, the discrete nonlinear
${\si}$-model will instead reside in a nontrivial disordered
phase~\cite{ZW180809394} that spontaneously breaks parts or all of the
generalized symmetry ${\cR}$~\cite{P230805730}. This causes the singularities
to become deconfined. Since the emergent symmetry has many different breaking
patterns, there can be many different disordered or partially disordered
phases. In what follows, we will focus on the maximally disordered phases.

In terms of the dual model, this nontrivial disordered
phase maps onto a trivial ${\mathbb{G}^{(d)}}$ symmetric phase. This
symmetry-breaking pattern is encoded by the magnetic boundary of the SymTO,
whose fusion ${d}$-category is ${\cR}$. We note that for ${\pi}$-finite
nonlinear ${\si}$-models, this nontrivial disordered phase is always gapped
since ${\cR}$ is a finite symmetry. For ${\pi}$-infinite nonlinear
${\si}$-models, since ${\cR}$ is a continuous symmetry, the disordered
phase will always have gapless excitations by Goldstone's theorem. 

In the remainder of this section, we will discuss some examples of disordered
phases for singularity-free discrete nonlinear ${\si}$-models.

\subsection{Singularity-free $\pi$-finite nonlinear ${\si}$-model} 

Consider a general singularity-free $\pi$-finite nonlinear ${\si}$-model in the
ordered phase. ${\cR}$ is a finite symmetry and does not include any ${d}$-form
symmetries since we assume the target space of the nonlinear ${\si}$-model is
connected. Therefore, the ${\cR}$ symmetry can always be spontaneously broken
(there are no Mermin-Wagner obstructions), and a nontrivial disordered phase
always exists.

Since ${\cR}$ is the magnetic symmetry of ${\mathbb{G}^{(d)}}$ higher gauge
theory, this disordered phase will correspond to the deconfined phase of
${\mathbb{G}^{(d)}}$ higher gauge theory. This will be a gapped phase where all
of the singularities are deconfined, and there will be ground state degeneracy
that depends on the topology of space. The ground states will be described by
an untwisted higher Dijkgraaf-Witten theory with ${d}$-group
${\mathbb{G}^{(d)}}$

\subsection{$T^2$ nonlinear ${\si}$-model}

Consider a singularity-free nonlinear ${\si}$-model in ${(d+1)}$ dimensional
spacetime whose target space ${K}$ is a 2-torus ${T^2 \cong S^1\times S^1}$.
Such a nonlinear ${\si}$-model describes a two-component superfluid, where two
${U(1)}$ symmetries are spontaneously broken.  However, to simplify our
discussion, let us explicitly break the two ${U(1)}$ symmetries.
Parametrizing the maps ${M_{d+1}\to T^2}$ by ${\phi_1(x),\phi_2(x)\in
2\pi\R/\Z}$, this nonlinear ${\si}$-model is described by the Lagrangian
\begin{align}
 \cL = |\prt_\mu \ee^{\ii \te_1(x)}|^2 +  |\prt_\mu \ee^{\ii \te_2(x)}|^2 + 
J_1 \cos(\te_1)
+J_2 \cos(\te_2)
\end{align}

The homotopy groups for this target space are
\begin{equation}
\pi_n(T^2) = \begin{cases}
\Z\times \Z,\quad\quad & n =1,\\
 0,\quad\quad & n \neq1.
 \end{cases}
\end{equation}
Therefore, there are two species of codimension 2
singularities---vortices---classified by ${\Z}$. From the previous section,
this implies there is an emergent ${U(1)\times U(1)}$ ${(d-1)}$-form symmetry.
Each of these ${U(1)}$ ${(d-1)}$-form symmetries are generated by a topological
defect line
\begin{equation}
D_{1,i}^{(\al)}(C) = \exp\left[\ii\al\int_C \frac{\dd\te_i}{2\pi}\right],
\end{equation}
where ${C}$ is a 1-cycle in spacetime. Indeed, in the absence of singularities,
${\dd(\dd\te_i) = 0}$ and ${D_{1,i}^{(\al)}(C)}$ depends only on the homology
class ${[C]}$.

There is also a ${U(1)}$ ${(d-2)}$-form symmetry arising from a composite
current ${\hstar j =
\frac{\dd\te_1}{2\pi}\wdg\frac{\dd\te_2}{2\pi}}$~\cite{B201200051}, where
${\hstar}$ is the Hodge star operator. It is generated by the topological
defect surface
\begin{equation}
D_2^{(\al)}(\Si) = \exp\left[\ii \al\int_\Si \frac{\dd\te_1}{2\pi}\wdg\frac{\dd\te_2}{2\pi}\right],
\end{equation}
where ${\Si}$ is a ${2}$-cycle in spacetime. Whereas ${D_{1,i}^{(\al)}(C)}$
detects species ${i}$ vorticies, ${D_2^{(\al)}(\Si)}$ detects Hopf-linked
vortices of species ${1}$ and ${2}$.

Does this ${T^2}$ nonlinear ${\si}$-model have a nontrivial disordered phase?
The ${U(1)\times U(1)}$ ${(d-1)}$-form symmetry cannot spontaneously break in
${(d+1)}$-dimensional spacetime~\cite{GW14125148, L180207747}. However, the
${U(1)}$ ${(d-2)}$-form symmetry can. Therefore, there is a nontrivial disordered phase driven by spontaneously breaking the ${U(1)}$
${(d-2)}$-form symmetry, which can be done by proliferating Hopf-linked
vortices.

Since the nontrivial disordered phase is a ${U(1)}$ ${(d-2)}$-form SSB phase,
it hosts a ${(d-2)}$-form gapless Goldstone boson ${a^{(d-2)}}$---an emergent
photon---and its long-wavelength properties are described by the effective
field theory
\begin{equation}\label{dm1FormMax}
S = \frac{1}{2e^2}\int_{M_{d+1}}\dd a^{(d-2)}\wdg\hstar \dd a^{(d-2)}.
\end{equation}
The ${U(1)}$ ${(d-2)}$-form symmetry's Noether current is given by ${\hstar j = \frac1{e^2}\hstar \dd a^{(d-2)}}$. Therefore, the Goldstone mode ${a^{(d-2)}}$ and the nonlinear ${\si}$-model fields ${\te_1}$ and ${\te_2}$ are related to one another by
\begin{equation}
\int_\Si \frac{\dd\te_1}{2\pi}\wdg \frac{\dd\te_2}{2\pi} = \frac{1}{e^2}\int_\Si \hstar\dd a^{(d-2)}.
\end{equation}
Roughly speaking, this means that the magnetic photon ${\hat{a}^{(1)}}$ of the Maxwell theory~\eqref{dm1FormMax} is related to ${\te_1}$ and ${\te_2}$ by ${\hat{a}^{(1)} = \frac{1}{2\pi}\te_1\dd\te_2}$.

\subsection{${3+1}$D singularity-free $S^2$ nonlinear ${\si}$-model}\label{S2Section}

We now consider the singularity-free nonlinear ${\si}$-model whose target space ${K}$ is the 2-sphere $S^2$. In the context of symmetry-breaking phases, this describes an isotropic antiferromagnetic where an ${SO(3)}$ symmetry is spontaneously broken down to ${SO(2)}$. The maps ${M_{d+1}\to S^2}$ can be parametrized by the unit vector ${\bm{n}\in\R^3}$, and the nonlinear ${\si}$-model is described by the Lagrangian 
\begin{align}
 \cL = \frac12(\prt_\nu \vc n)^2.
\end{align}

The $S^2$ nonlinear $\si$-model is historically studied using the large $N$ limit. The three-dimensional unit vector ${\bm{n}}$ is replaced with an ${N+1}$ dimensional unit vector, causing the target space to become the $N$-sphere $S^N$~\cite{GJ7810}. This large $N$ analysis predicts that the only disordered state is a gapped product state. This is consistent with the result of this framework presented in this paper: since ${\pi_n(S^N) = 0}$ for ${n=0,1\cdots,N-1}$, the $S^N$ nonlinear $\si$-model has no generalized symmetry for large $N$ and its disordered state is the gapped product state.\footnote{Another large $N$ limit is to replace the $S^2$ target space with $\C P^{N-1}$ since ${S^2 = \C P^1}$~\cite{DDL7863,W7985}. Studying this large $N$ limit shows that the disordered state in 3+1 dimensional spacetime is described by a gapless $U(1)$ gauge theory. This is consistent with the result of this paper. Since ${\pi_2(\mathbb{C}P^{N-1}) = \Z}$ and ${\pi_n(\mathbb{C}P^{N-1}) = 0}$ for ${n=0,1,3,4,5, \cdots, 2N-2}$, the $\C P^{N-1}$ nonlinear $\si$-model has a $U(1)$ 1-form symmetry.  Its disordered state spontaneously breaks this $U(1)$ 1-form symmetry, which produces a state with gapless mode described by a $U(1)$ gauge field. } In what follows, we show that this large $N$ analysis of the $S^2$ nonlinear $\si$-model misses an interesting disordered phase described by massless axion electrodynamics.

The first six homotopy groups of ${S^2}$ are 
\begin{align}
\pi_1(S^2) &\simeq 0, & \pi_2(S^2) &\simeq \Z, & \pi_3(S^2) &\simeq \Z,
\nonumber \\
\pi_4(S^2) &\simeq \Z_2, & \pi_5(S^2) &\simeq \Z_2, & \pi_6(S^2) &\simeq \Z_{12}.
\end{align}
Since the homotopy groups of ${S^2}$ are so rich, so is the ${d}$-group
${\mathbb{G}^{(d)}}$ (see Eq.~\eqref{dGrouphomotopyGroup}), and thus the
generalized symmetries of the ${S^2}$ nonlinear ${\si}$-model, will also be
interesting and complex. Since ${\pi_1(S^2)}$ is trivial, the group
homomorphisms ${\al_n\colon \pi_1(S^2)\to \text{Aut}(\pi_n(S^2))}$ are all
trivial. However, the Postnikov invariants of the classifying space of
${\mathbb{G}^{(d)}}$ will generally be nontrivial, causing the model's
generalized symmetries to be nontrivial. 

In what follows, we will exclusively work in ${(3+1)}$D spacetime as this is
the simplest nontrivial case. The higher group ${\mathbb{G}^{(d)}}$ is then a
${3}$-group whose classifying space is (see Appendix~\ref{Postnikov})
\begin{equation}\label{3grpS2}
B\mathbb{G}^{(3)} = S^2_{\tau\leq 3} = \cB_{c_4}( 0,\Z,\Z).
\end{equation}
This 3-group is an extension of a ${\Z}$ 1-form symmetry group ${\Z^{(1)}}$ by a ${\Z}$ 2-form symmetry group ${\Z^{(2)}}$,
\begin{equation}\label{G3SES}
0\to \Z^{(2)}\to  \mathbb{G}^{(3)}\to \Z^{(1)}\to 0,
\end{equation}
and it is characterized by the cohomology class ${[c_4] \in H^4(K(\Z,2),\Z)}$. Such extensions of higher-form symmetry groups were discussed in \Rf{T171209542}. 

In appendix Sec.~\ref{hgroup}, we have introduced canonical ${2}$ and ${3}$
cochains ${x^{2}}$ and ${x^{3}}$, respectively, on $\cB_{c_4}( 0,\Z,\Z)$, which
characterize the homotopy type of $\cB_{c_4}( 0,\Z,\Z)$ and the above
extension, via the following relation
\begin{equation}\label{S23grpcocycleconds}
\dd x^2 = 0,\quad\quad \dd x^3 = c_4(x^2),
\end{equation}
where ${c_4}$ is a cocycle representative of ${[c_4]}$.

So what is the cocycle $c_4$? What we show next is that
\begin{equation}\label{c4forS2}
c_4(x^2) = x^2\smile x^2 \equiv \Sq_\Z^2 (x^2).
\end{equation}
Therefore, ${\mathbb{G}^{(3)}}$ is a discrete, nonfinite ${3}$-group with a nontrivial Postnikov invariant. Following the discussion from Sec.~\ref{piinf}, since the cup product in~\eqref{c4forS2} is not a stable cohomology operation, there is no dual ${3}$-group to ${\mathbb{G}^{(3)}}$. Consequently, the generalized symmetries of the ${S^2}$ nonlinear ${\si}$-model is a non-invertible symmetry. This agrees with the result from \Rf{CT221013780}.

Let us now justify Eq.~\eqref{c4forS2}. This is a standard argument in algebraic topology. We refer the reader to Appendix~\ref{QuickPostnikov} for a quick, less formal argument. We start with the Hopf map ${\eta \colon S^3 \to S^2}$. This is, by definition, the attaching map for the $4$-cell in $\C P^2$. We can construct a commutative diagram
\[\xymatrix{ S^3 \ar[r]^\eta \ar[d]^-{a^3}  & S^2 \ar[r]^-{\subset} \ar[d]^-{a^3} & \C P^2 \ar[d]^-{a^3}   \ar[r]^-q &\C P^2/S^2 \cong S^4 \ar[d]^-{a^4}  \\
K(\Z,3) \ar[r]^{i_3}  & S^2_{\tau\leq 3} \ar[r]^-{p^3} & K(\Z,2)  \ar[r]^-{k^4} & K(\Z,4)
}\]
where $q$ is the quotient map. The three maps labeled $a^3$ are the maps in the Postnikov towers for the truncations ${\tau \leq 3}$ for each respective space, while $a^4$ is the map in the Postnikov tower for $S^4$ corresponding to the truncation ${\tau\leq 4}$. That this diagram commutes is implied in the construction of $k^4$, as explained in the proof of the only Theorem of Ch.22 \S4 in \Rf{MayConcise}. Some of the maps in our diagram are cohomology classes because their targets are Eilenberg-MacLane spaces. For example, ${[a^3] \in H^2(\C P^2;\Z)}$ and ${[a^4q] \in H^4(\C P^2;\Z)}$. The fact that $\eta$ has Hopf invariant one is equivalent to the equation ${[a^4q] = [(a^3)^2]}$. But since $a^3$ induces an isomorphism ${H^4(K(\Z,2),\Z) \cong H^4(\C P^2,\Z)}$ and ${[a^4q ] =[k^4a^3]}$, we see that $[k^4]$ must be the square of the fundamental class in ${H^2(K(\Z,2),\Z)}$, so that ${[k^4]=\Sq_\Z^2}$. Since the homotopy class of ${k^4}$ is classified by the cohomology class of ${c_4}$, this proves Eq.~\eqref{c4forS2}. 

Let us now consider the nontrivial disordered phase arising from spontaneously breaking this generalized symmetry. It is usually believed that the disordered phase of such a $S^2$ nonlinear ${\si}$-model is a trivial phase described by a product state.  But such a result is correct only if the singularities are proliferated in the disordered phase. Otherwise, the disordered phase is nontrivial. As we will now show, the disordered phase of the singularity-free ${S^2}$ nonlinear ${\si}$-model has two gapless modes---a gapless scalar and an emergent photon---and is described by massless axion electrodynamics.

We show this by constructing a low-energy effective field theory of 3+1D $S^2$
non-linear $\si$-model describing the nontrivial disordered phase.  We first
replace the target space $S^2$ by its 3rd Postnikov stage $S^2_{\tau\leq 3} =
\cB_{c_4}( 0,\Z,\Z)$.  By replacing $S^2$ with $S^2_{\tau\leq 3}$, we describe
the singularity-free disordered phase of the non-linear $\si$ model in which
the $3\text{-}\mathsf{Rep}(\mathbb{G}^{(3)})$ symmetry is maximally
spontaneously broken.  The non-linear $\si$-model is now described by the
following path integral
\begin{align}
 Z = \sum_{\vphi} e^{-S(\vphi)}
\end{align}
where $\vphi: M_{3+1} \to S^2_{\tau\leq 3}$ are singularity-free (\ie continuous)
map from the spacetime $M_{3+1}$ to the target space $S^2_{\tau\leq 3}$. 

The canonical ${2}$ and ${3}$ cochains ${x^{2}}$ and ${x^{3}}$ on
$S^2_{\tau\leq 3}$, characterizing the homotopy type of $S^2_{\tau\leq 3}$,
have the following relation
\begin{equation}
\dd x^2 = 0,\quad\quad \dd x^3 = x^2\smile x^2.
\end{equation}
Let $\frac{F}{2\pi} = \vphi^* x^2$ be the 2-form field on $M_{3+1}$ which is
the pull back of $x^2$ on $S^2_{\tau\leq 3}$.  Similarly, let $\frac{H}{2\pi} =
\vphi^* x^3$ be the pull back of $x^3$ on $S^2_{\tau\leq 3}$.  We now use
the fields $(F,H)$ to represent the maps $\vphi$, and consider
the path integral as a summation over fields $(F,H)$ (with constraints),
instead of over the maps $\vphi$:
\begin{align}
\label{ZFH}
 Z &= \sum_{F,H} e^{-S(F,H)},
\nonumber\\
 \dd F &=0, \ \ \dd H = (2\pi)^{-1}F\wedge F . 
\end{align}
The simplest effective action is
\begin{align}
\label{SAB}
 S(F,H) = \int_{M_{3+1}} \frac{1}{2e^2} |F|^2 + \frac{1}{4\pi v^2 } 
|H|^2,
\end{align}
where, for instance, ${|F|^2 \equiv F\wdg\hstar F}$.

Since ${\dd F=0}$, we can introduce a $U(1)$ gauge field $A$ to describe $F$ via ${F=\dd A}$.
In fact, such a gauge field $A$ appears in the
${\mathbb{C}P^1}$ representation of the $S^2$ non-linear
$\si$-model~\cite{Pol87} which has a compact ${U(1)}$ gauge field ${A = A_\mu
\dd x^\mu}$ and a two-component unit spinor field ${z = (z_1,z_2)\in \C^2}$
related to ${\vc n}$ by the Hopf map
\begin{align}
 \vc n = z^\dag \vc \si z,
\end{align}
where ${\vc \si = (\si^1,\si^2,\si^3)}$ are the Pauli matrices. At the level of the equations of motion---the level of local physics and trivial ${U(1)}$ bundles---${A}$ is related to ${z}$ by
\begin{align}
 A_\mu = \ii z^\dag \prt_\mu  z .
\end{align}
In particular, 
the homotopy classes ${\pi_2(S^2) \simeq \Z}$ are characterized
by the Chern number of the $U(1)$-bundle,
\begin{align}\label{CP1pi2}
 \int_{S^2} \frac{1}{2\pi} \dd A \in \pi_2(S^2)\simeq\Z,
\end{align}
and the homotopy classes ${\pi_3(S^2)}$ are characterized by the Hopf invariant, which can be expressed in terms of $A_\mu$ as~\cite{Ah0005150}
\begin{align}\label{CP1pi3}
 \int_{S^3} \frac{1}{4\pi^2}  A\wdg\dd A \in \pi_3(S^2)\simeq\Z.
\end{align}

In light of Eqs.~\eqref{ZFH}, \eqref{CP1pi2} and \eqref{CP1pi3}, we construct the low-energy field theory using dynamical ${1}$-form and ${2}$-form ${U(1)}$ gauge fields ${A}$ and ${B}$, respectively. The gauge field ${A}$ is the same that appears in the ${\mathbb{C}P^1}$ presentation. The gauge field ${B}$ is introduced to remedy the fact that the integrand in Eq.~\eqref{CP1pi3} is not gauge invariant. Using ${A}$ and ${B}$, we construct the gauge invariant fields $F$ and $H$, which are expressed as
\begin{align}
F &= \dd A,\\
 H &=  \frac{1}{2\pi}A\wdg\dd A + \dd B,
\end{align}
and correctly satisfy the constraints
$ \dd F =0, \ \dd H = (2\pi)^{-1}F\wedge F$.
The periods of ${\frac1{2\pi}F}$ and ${\frac1{2\pi}H}$ are in ${\pi_2(S^2)}$ and ${\pi_3(S^2)}$, respectively. Because only ${F}$ and ${H}$ are physical, ${A}$ and $B$ have the gauge redundancy
\begin{align}
 A &\to A+\dd \al,\\
 B &\to B+\dd \bt - \frac{1}{2\pi} \al \dd A.
\end{align}
Interestingly, this gauge redundancy has the same form as the continuous 2-group gauge redundancy discussed in \Rf{BH180309336}. 

We can dualize the ${2}$-form field ${B}$ to get the action Eq.~\ref{SAB} into
a more standard form. Integrating out ${H}$ directly while including a Lagrange
multiplier field ${\phi}$ to ensure its modified Bianchi identity
\begin{equation}
\frac{\dd H}{2\pi} = \frac{F}{2\pi}\wdg \frac{F}{2\pi}
\end{equation}
is satisfied, we find the dual action
\begin{align}
\label{SAphi}
 S =
\int_{M_{3+1}} \frac{1}{2e^2} |F|^2
+ \frac{v^2}{2 } |\dd\phi|^2 + \frac{1}{4\pi^2 }\phi~ F\wdg F.
\end{align}
This is nothing but axion electrodynamics with the photon field ${A}$ and massless axion field ${\phi}$. The masslessness of these is protected by the spontaneously broken generalized symmetry. We remark that the axion coupling ${\phi~ F\wdg F}$ arises as a direct consequence of the nontrivial Postnikov class ${[c_4]}$~\eqref{c4forS2}. 

The generalized symmetries of ${(3+1)}$D axion electrodynamics have been intensely studied and shown to be a rich structure of both invertible and non-invertible 0-form and higher-form symmetries~\cite{HNY200612532, HNY200914368, BC201109600, CLS221204499, Y221205001}. It is challenging to compare all of the non-invertible symmetries of the $S^2$ non-linear $\si$-model with those of massless axion electrodynamics since the latter will include emergent symmetries. The $3\text{-}\mathcal{R}\mathrm{ep}(\mathbb{G}^{(3)})$ symmetry we have based of discussion around includes a $\Q/\Z$ non-invertible symmetry~\cite{CT221013780}. Our calculation here shows that spontaneously breaking this symmetry gives rise to massless axion-electrodynamics. The $\Q/\Z$ non-invertible symmetry of $3\text{-}\mathcal{R}\mathrm{ep}(\mathbb{G}^{(3)})$ matches the $\Q/\Z$ non-invertible symmetry of massless axion-electrodynamics discussed in \Rf{CLS221204499}.

This result is particularly interesting from a condensed matter physics point of view, where axion electrodynamics has also been found to emerge in theories describing topological insulators~\cite{LWQ09081537}, quantum spin liquds~\cite{PCC210906890}, topological superconductors~\cite{QWZ12061407}, and Weyl semimetals~\cite{WZ12075234}. From this point of view, we find that starting in an isotropic antiferromagnet---where ${SO(3)}$ is spontaneously broken to ${SO(2)}$---in ${(3+1)}$D and disordering without proliferating defects induces a transition into a phase described by axion electrodynamics. An interesting and important follow-up direction to this result is understanding in a microscopic spin Hamiltonian what terms would drive the transition from an anti-ferromagnet phase to this axion electrodynamics phase.

\section{Conclusions}

In this paper, we have investigated the phases of nonlinear ${\si}$-models from the point of view of generalized symmetries. It is usually believed that the disordered phase of nonlinear ${\si}$-models is a trivial phase described by a product state.  But such a result is correct only if the singularities are proliferated in a disordered phase. Otherwise, as emphasized here, the disordered phase is nontrivial and contains spontaneously broken emergent generalized symmetries.

The emergent generalized symmetries discussed in Secs.~\ref{emsymm} and~\ref{piinf} therefore offer a way to classify the disordered phases of nonlinear ${\si}$-models. The trivial disordered phase is one where these symmetries can no longer emerge. The nontrivial disordered phases, which we studied here, occur when the generalized symmetries still emerge and are classified by their spontaneous symmetry-breaking patterns. We discussed examples of this in Sec.~\ref{examples}.

\let\oldaddcontentsline\addcontentsline
\renewcommand{\addcontentsline}[3]{}
\section*{Acknowledgements}
\let\addcontentsline\oldaddcontentsline

We are grateful for helpful discussion with Hank Chen, Theo Johnson-Freyd, and Ryan Thorngren.
S.D.P. is supported by the National Science Foundation Graduate Research Fellowship under Grant No. 2141064 and by the Henry W. Kendall Fellowship.
This work is partially supported by NSF DMR-2022428, DFG ZH 274/3-1
and by the Simons Collaboration on Ultra-Quantum Matter, which is a grant from
the Simons Foundation (651446, XGW). This material is based upon work supported by the National Science Foundation under Grant No. DMS 2143811 (A.B.).

\appendix

\section{Classifying spaces of higher groups as simplicial sets}\label{hgroup}

In this appendix, we will describe an explicit construction for
the nerve of an ${n}$-group as a simplicial set (see Appendix~\ref{SSet} for background on simplicial sets). The realization of the nerve of an $n$-group is the classifying space of the $n$-group. We remark that, in physics, the classifying space of an ${n}$-group
describes the gauge fields, dynamical or background, associated with an
${n}$-group gauge redundancy. We will first describe the construction explicitly and then show why it is the nerve of an ${n}$-group.

We denote the simplicial set for the nerve of our $n$-group by
\begin{equation}\label{eq:Bdef}
\cB= \cB_{\alpha_2, c_3;\alpha_3, c_4 ; \cdots ; \alpha_n, c_{n+1}} (G,A_2, \cdots, A_{n}).
\end{equation}
The data in the notation of  \eqref{eq:Bdef} consists of:
\begin{enumerate}
\item A collection of discrete\footnote{Although we restrict ourselves to discrete groups, classifying spaces and nerves of general higher groups can depend on continuous groups.} groups ${G, A_2, \cdots, A_n}$, with the condition that the $A_i$'s are abelian groups. Topologically, these represent the homotopy groups ${\pi_1,\ldots,\pi_n}$ of a space $K$. 
\item  For each ${j\in\{2,3,\cdots, n\}}$, a group homomorphism
\begin{equation}
\al_j\colon G\to \text{Aut}(A_j),
\end{equation}
where ${\text{Aut}(A_j)}$ is the group of automorphisms of ${A_j}$. These describe an action of ${G}$ on ${A_j}$ which, topologically, corresponds to the action of the fundamental group of $K$ on its higher homotopy groups. We will let $A_j^{\alpha_j}$ be the group $A_j$ equipped with the action of $G$ specified by $\alpha_j$.
\item Inductively defined twisted cocycles $c_3,\ldots, c_{n+1}$ representing cohomology classes
\[[c_{k+1}]\in H^{k+1}(\cB_{\al_2,c_3; \cdots; \al_{k-1},c_{k}} (G,
A_2, \dots, A_{k-1}), A_k^{\alpha_{k}})\]
for ${k=2, \cdots, n}$.
Topologically, these are related to the  Postnikov $k$-invariants of $K$ and generalizations thereof in certain twisted cohomology groups.
\end{enumerate}
These simplicial sets should model all connected homotopy $n$-types, i.e., homotopy types of connected topological spaces with only finitely many non-zero homotopy groups.\footnote{In fact, a proof that we get all homotopy $n$-types should be straightforward to assemble from the mathematics literature by combining \Rf{Droretal}, Theorem 1.4, \Rf{may}, Proposition 25.2 and \Rf{Gitler}, Theorem 7.18. Note \Rf{Robinson}, Theorem 3.4 gives a treatment using topological spaces instead of simplicial sets. When $G$ is abelian and the $\alpha_k$'s are all trivial, the cohomology classes of the $c_{k+1}$'s are the Postnikov $k$-invariants of the space $K$. See \Rf{may}, Theorem 25.7, Remark 25.9 (2), as well as the discussion in Appendix~\ref{Postnikov}. More generally, the $\alpha$'s give rise to local coefficient systems and the $c_{k+1}$ are twisted cohomology classes. See also \S3.2.2 of \Rf{thorngrenThesis} and \S3.3 of \Rf{BaezShulman} useful discussions.}

We elaborate on the relationship between $\cB$ and $K$. Given a simplicial set, there is a procedure for producing a topological space $K$ by using the data of $\cB$ to glue together simplices. Then $K$ is called the \emph{realization} of $\cB$ and is often denoted by ${K=|\cB|}$ (see \Rf{may} \S14).  Above, we called $\cB$ a \emph{simplicial-set triangulation} of our target space $K$ even though it only gives us a CW-structure, which might not be a triangulation in the strict sense of the term.

\bigskip 

We begin our description of $\cB$. It is a simplicial set, so is determined by a set
of vertices $[\cB]_0$, a set of links $[\cB]_1$ (i.e., $1$-simplices), a set of triangles  $[\cB]_2$ (i.e., $2$-simplices),
etc.  They are formally related to each other by face maps
\begin{equation}
\label{eq:nerveB}
\xymatrix{ 
[\cB]_0 & 
[\cB]_1 \ar@<-1ex>[l]_{d_0, d_1}\ar[l] & 
[\cB]_2 \ar@<-1ex>_{d_0, d_1 , d_2}[l] \ar@<1ex>[l] \ar[l] & 
[\cB]_3 \ar@<-1ex>[l]_{d_0, ..., d_3} \ar@<1ex>[l]_{\cdot} & 
[\cB]_4 \ar@<-1ex>[l]_{d_0, ..., d_4} \ar@<1ex>[l]_{\cdot}  ,
}
\end{equation}
where $d_i$ are the face maps, describing how an $n$-simplex should be attached to
the ${(n-1)}$-simplices.\footnote{There are also degeneracy maps which are discussed later.}

\begin{figure}[t!]
\centering
    \includegraphics[width=.48\textwidth]{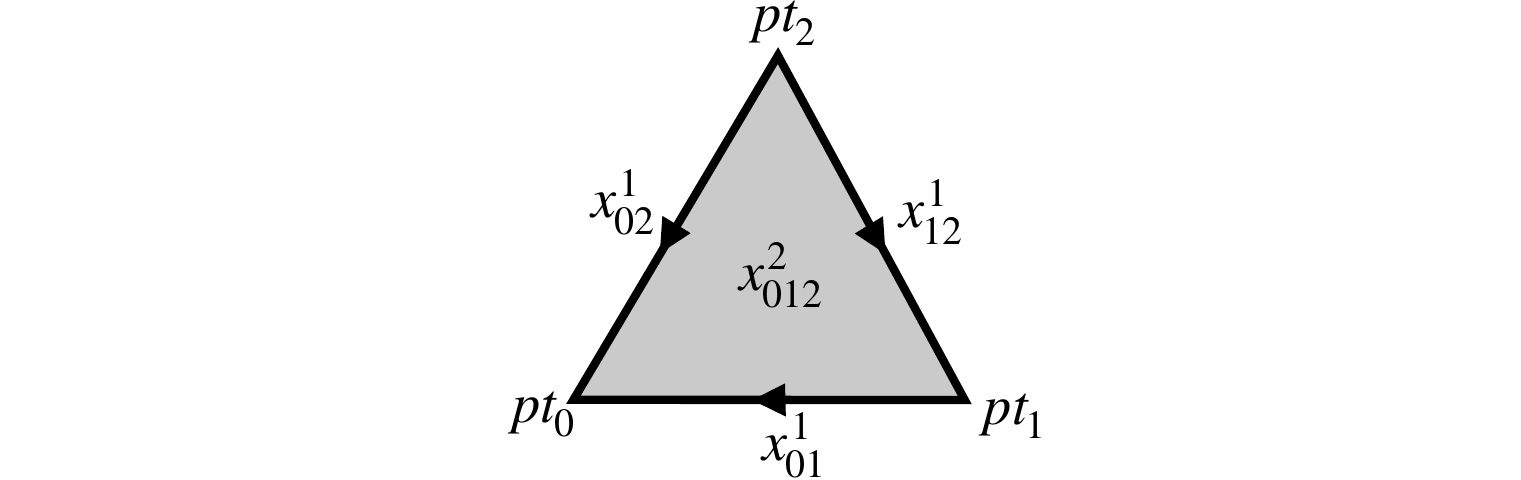}
    \caption{A triangle or $2$-simplex in $[\cB]_2$. The labels are subject to the conditions ${x_{pq}^1\in G}$ with ${x^1_{12}  x^1_{01}=x^1_{02}}$ and ${x_{012}^2\in A_2}$. The vertices ${pt_0,pt_1,pt_2}$ are all identified with the single $0$-vertex $pt$.} \label{fig:tetr2} 
\end{figure}

The nerve $\cB$ of our $n$-group has a single vertex ${[\cB]_0 = \{pt\}}$. The set of links $[\cB]_1$ is the group $G$. Geometrically, one thinks of  ${[\cB]_1=G}$ as a set of labels for a collection of links that each start and end
at the vertex $pt$. 
A $2$-simplex, or triangle in $[\cB]_2$ will be labeled by group elements ${x^1_{01},x^1_{12},x^1_{02}}$ in $G$, thought of as decorating the three edges of the triangle, and an
additional label $x^2_{012}$ which takes values in
$A_2$, and is a label for the unique face of the triangle.   See Figure~\ref{fig:tetr2}. So, the elements of   $[\cB]_2$ are tuples ${(x^1_{01},x^1_{12},x^1_{02};
x^2_{012}) \in G^3 \times A_2}$. However, not all such labels are admissible as elements of  $[\cB]_2$, only those that satisfy certain conditions we will define next.

We  introduce the compact notation 
\begin{align} 
\label{s012}
s[012] = (x^1_{01},x^1_{02}, x^1_{12}; x^2_{012})  
\end{align} 
to denote a decorated triangle.  With this convention, for a labeling of the simplices, the face maps ${d_m: [\cB]_2
\to [\cB]_1}$ can be expressed simply as
\begin{align}
 d_0 (x^1_{01},x^1_{02}, x^1_{12};x^2_{012}) &= x^1_{12},
\nonumber\\
 d_1 (x^1_{01},x^1_{02}, x^1_{12};x^2_{012}) &= x^1_{02},
\nonumber\\
 d_2 (x^1_{01},x^1_{02}, x^1_{12};x^2_{012}) &= x^1_{01}.
\end{align}
A decorated triangle ${s[012]=(x^1_{01},x^1_{02}, x^1_{12}; x^2_{012})}$ is in $[\cB]_2$ if and only if 
\[ d_0(s[012])d_1(s[012])^{-1}d_2(s[012]) =x^1_{12} (x^1_{02})^{-1} x^1_{01}=1\in G.\]

More generally, a $d$-simplex in $[\cB]_d$ will correspond to a tuple ${s[0\cdots d]=(x^1_{pq};x^2_{pqr};\cdots; x^d_{0\cdots d})}$, 
\begin{equation}\label{eq:labels}
s[0\cdots d] \in G^{\binom{d+1}{2}}\times A_2^{\binom{d+1}{3}}\cdots \times A_d^{\binom{d+1}{d+1}}.
\end{equation}
with the entries of ${s[0\cdots d]}$ subject to inductively defined restrictions that we will specify later. 
When ${d>n}$, $A_d$ denotes the trivial group.

 Geometrically, the labels $x_{pq}^1$ decorate the links of a geometric $d$-simplex, the $x_{pqr}^2$ decorate the triangles, etc., and the $x_{0\cdots d}^d$ decorates the solid interior of the $d$-simplex.

The face map ${d_m\colon [\cB]_d
\to [\cB]_{d-1}}$ is given by 
\begin{align}
\label{dms}
 d_m( s[0\cdots d] )= s[0\cdots \hat m \cdots d],
\end{align}
where $\hat m$ means that we omit the $m$th index. Notice that the face map $d_m$ is obtained by applying the co-face map $d^m$ of \eqref{eq:coface} below to the indices of the labels.\footnote{Similarly, the degeneracy ${s_m \colon [\cB]_d \to [\cB]_{d+1}}$ is obtained by applying the co-degeneracy $s^m$ of \eqref{eq:codegen} to the indices of the labels, with the understanding that a label with a repeated index is the identity element. So, ${s_m(s[0\cdots d])= (y_{p_0p_1}^1;\cdots ; y_{p_0\cdots p_d}^{d} ; 0)}$, where ${y_{p_0\cdots p_j}^j = x_{s^m(p_0) \cdots s^m(p^j)}^{j}}$. For example, ${s_0(x_{01}^1) = (1, x_{01}^1, x_{01}^1; 0)}$ and ${s_1(x_{01}^1) = (x_{01}^1,x_{01}^1 ,1; 0)}$.}
Each $d$-simplex has $\binom{d+1}{d-j}$ $j$-faces given by
\begin{align*}
d_{q_1}d_{q_{2}} \cdots d_{q_{j-d}}s[0\cdots d]
&=s[0\cdots \hat{q}_1 \cdots \hat{q}_{j-d} \cdots d].
\end{align*}
for ${0\leq q_1< \ldots< q_{d-j} \leq d}$. For example, a tetrahedron has 4 triangles (2-faces) and 6 links (1-faces). 

To explain the additional restrictions on ${s[0\cdots d]}$, we introduce the canonical cochains on $\cB$. The canonical $G$-valued 1-cochain $x^1$ is given by
its evaluation on 1-simplices ${s[01]=x^1_{01}}$:
\begin{align}
 \<x^1, x^1_{01}\> = x^1_{01},\ \ \ \ x^1_{01}\in G.
\end{align}
The canonical $A_2$-valued 2-cochain $x^2$ is given by its evaluation on 2-simplices ${s[012]=(x^1_{01},x^1_{12},x^1_{02};x^2_{012})}$:
\begin{align}
\<x^2,(x^1_{01},x^1_{12},x^1_{02};x^2_{012})\> = x^2_{012},\ \ \ \ 
x^2_{012}\in A_2, 
\end{align}which is simply the projection to $A_2$. Thus, in general, the canonical $A_d$-valued $d$-cochain $x^d$ is defined as the function
\[x^d \colon G^{\binom{d+1}{2}}\times A_2^{\binom{d+1}{3}}\cdots \times A_d^{\binom{d+1}{d+1}} \to A_d\]
which projects onto the last coordinate.
We also let ${\dd x^1  \colon G^{\binom{3}{2}} \times A_2^{\binom{3}{3}} \to G}$ be given by
\begin{align*}
\dd x^1(s[012])& = x^1(s[12])  x^1(s[02])^{-1} x^1(s[01]) 
\end{align*}
and
${\dd_{\alpha_k}x^k  \colon G^{\binom{k+2}{2}} \times A_2^{\binom{k+2}{3}} \times \cdots \times A_{k+1}^{\binom{k+2}{k+2}} \to A_{k}}$
by
\begin{equation}
\begin{aligned}
 \dd_{\alpha_k}x^k(s[0\cdots (k+1)])
 &= \alpha_k(x_{01}^1)x^k(s[1\cdots (k+1)])  \\
  &   + \sum_{i=1}^k(-1)^ix^k(s[0\cdots\hat{i} \cdots (k+1)]).
\end{aligned}
\end{equation}
Note that these formulas are obtained by composing the fundamental cochains with alternating sums of face maps, using the $\alpha$'s to twist the first term in the sum.

We've already specified that the simplices are tuples ${s[0\cdots d]}$ in ${G^{\binom{d+1}{2}}\times A_2^{\binom{d+1}{3}}\cdots \times A_d^{\binom{d+1}{d+1}}}$ and we have given the face and degeneracy maps. It remains to describe the restrictions on the tuples. These are given inductively, so we will let
\begin{equation}\label{eq:Bk}
\cB^k:= \cB_{\alpha_2,c_3;\cdots;\alpha_k, c_{k+1}}(G,A_1,\cdots, A_k).
\end{equation}

We start with ${k=1}$ and complete our definition of $\cB^1$. We have already described ${[\cB]_0, [\cB]_1, [\cB]_2}$. More generally, a label ${s[0\cdots d] \in G^{\binom{d+1}{2}}}$ is in $[\cB^1]_d$ if and only if every $2$-face of ${s[0\cdots d]}$ satisfies the equation ${\dd x^1=1}$. That is,
${x^1_{pq}(x^1_{pr})^{-1} x^1_{qr}=1}$
for all ${p,q,r \in \{0,\ldots, d\}}$ for the ${x^1_{..}}$ labels in ${s[0\cdots d]}$. We see that $[\cB^1]_d$ is in bijection with $G^{d}$, where we send ${(x_{ij}^1)}$ to ${(g_1,\cdots, g_d)}$ with ${g_i = x_{(i-1)i}}$. In fact, $\cB^1$ is one of the standard models for the classifying space $BG$. 

Given ${\alpha_2 \colon G \to \mathrm{Aut}(A_2)}$, we define a cochain complex ${C^*(\cB^1 ;A_2^{\alpha_2})}$ as follows. The $k$-cochains are functions ${\varphi \colon [\cB^1]_k \to A_2}$ from the $k$-simplices of $\cB^1$ to the abelian group $A_2$. The differential 
\[\dd_{\alpha_2} \colon  C^{k-1}(\cB^1 ;A_2^{\alpha_2}) \to C^k(\cB^1 ;A_2^{\alpha_2}) \] 
is given by
\begin{align*}
\dd_{\alpha_2}\varphi(s[0\cdots k]) = & \ \alpha_2(x_{01}^1)  \varphi(s[1\cdots k]) \\
&+ \sum_{i=1}^k (-1)^i \varphi(s[0\cdots \hat{i}\cdots k] ).
\end{align*}

We let ${c_3 \in Z^3(\cB^1 ;A_2^{\alpha_2})}$ be a normalized cocycle.\footnote{In terms of the simplicial set $\cB^1$, normalization means that the function is zero on the image of the degeneracy maps $s_m$. If we identify ${Z^3(\cB^1,A_2^{\alpha_2})}$ with the group cocycles ${Z^3(G,A_2^{\alpha_2})}$, normalization means that the $c_3$ is zero when any of its input is the identity element of $G$.}
Given such a choice of $c_3$, we can construct $\cB^2$
 as follows. 
We let $[\cB^2]_d$ be the set of labels ${s[0\cdots d]}$ as in \eqref{eq:labels} whose $2$-faces and $3$-faces satisfy, respectively,
\begin{align*}
\dd x^1=1 \quad  \quad \text{and} \quad  \quad 
\dd_{\alpha_2}x^2=c_3.
\end{align*}
Here, to evaluate $c_3$, it is implicit that we forget the $A_2$ labels in ${s[0\cdots d]}$ and only remember the $G$-labels. 

Suppose that we have defined $\cB^{n-1}$.
Given $\alpha_n$, we can form a cochain complex  ${C^*(\cB^{n-1}; A_n^{\alpha_n})}$ whose $k$-cochains are functions ${\varphi \colon [\cB^{n-1}]_k \to A_n}$ on the $k$-simplices $ [\cB^{n-1}]_k $ of $\cB^{n-1}$. The differential is
\begin{align*}
\dd_{\alpha_n}\varphi(s[0\cdots k]) = & \alpha_n(x_{01}^1)  \varphi(s[1\cdots k]) \\
&+ \sum_{i=1}^k (-1)^i \varphi(s[0\cdots \hat{i}\cdots k] ).
\end{align*}
After choosing a normalized ${n+1}$-cocycle ${c_{n+1} \in Z^{n+1}(\cB^{n-1} ;A_n^{\alpha_n})}$, we define $\cB^{n}$
to have $d$-simplices $[\cB^n]_d$ those elements ${s[0\cdots d]}$ whose faces satisfy the conditions
\begin{align*}
\dd x^1&=1  \\
\dd_{\alpha_j}x^j &= c_{j+1} & 2\leq j\leq d-1.
\end{align*}
Again, it is implicit that to evaluate $c_{j+1}$, we only remember the ${G,A_2, \ldots, A_{j-1}}$ labels, and if ${d>n}$ the range on the second condition stops at $n$.

In summary,  the simplicial set 
\[\cB=\cB^n =\cB_{\al_2,c_3; \dots;
\al_n, c_{n+1}}(G, A_2, \dots, A_n)\] 
has $d$-simplices described by 
\begin{align}
\label{Xd}
\begin{split}
[\cB]_d & :=\{ s[0\dots  d]=(x^1_{01}, x^1_{02}, \dots, x^1_{(d-1)d} ; \\
& x^2_{012}, \dots, x^2_{(d-2)(d-1)d}; \dots; x^d_{0\dots d} ) | \\
 &x^1_{..}\in G,\ \   x^j_{..} \in A_j, \\
& \dd_{\alpha_{j}} x^j =  c_{j+1}(x^1; x^2; \dots;
  x^{j-1}), \\ 
&
\forall j=2, 3, \dots, d-1, \text{ and } \;
\dd x^1= 1 \}
\end{split}
\end{align}

It still remains to argue that the simplicial set ${\cB} $ we constructed corresponds to the nerve of an $n$-group. To do this, we need to verify that various Kan conditions are satisfied. These conditions are described in  Appendix \ref{SSet}.
Specifically, for all ${0\leq j\leq m}$, our simplicial set $\cB$ must satisfy ${\Kan(m, j)}$ for all ${m\geq 1}$ and ${\Kan!(m, j)}$ for all ${m\geq n+1}$.
The Kan conditions refer to the map
\[[\cB]_m \to \Lambda_{j}^m(\cB)\]
from $m$-simplices to $m$-horns defined in \eqref{eq:Kan_arrow}. See also Figure~\ref{fig:bimodule_A1} for intuition on horns. The condition ${\Kan(m, j)}$ is satisfied if this map is surjective, while ${\Kan!(m, j)}$ is satisfied if it is bijective. Colloquially, ${\Kan(m, j)}$ holds if any horn $\Lambda^m_j$ can be completed to an $m$-simplex $\Delta^m$, and ${\Kan!(m, j)}$ holds if there is a unique choice of $m$-simplex for completing each horn.

Notice that the horn space $\Lambda_{j}^m(\cB)$ has the
same ${(m-2)}$-skeleton as $\cB$, thus to verify the Kan condition
${\Kan(m, j)}$, we only need to take care of the missing ${(m-1)}$-face and $m$-face in the horn.  

An $m$-simplex of $\cB$ for $m \ge n+1$ is uniquely determined by its $n$-faces, so it is clear that ${\Kan!(m,
j)}$ is satisfied whenever ${m\geq n+2}$. Indeed, specifying a horn $\Lambda^m_j$ specifies its $n$-faces and, thus, there is a unique $m$-simplex with these specified $n$-faces.

Next, the conditions ${\Kan!(n+1, j)}$ are satisfied for ${0\le j \le n+1}$ because the
 equation
\begin{equation}\label{eq:last-one}
\dd_{\alpha_n} x^n = c_{n+1} (x^1; x^2; \dots; x^{n-1}),
\end{equation}
implies that as long as we know any ${n+1}$ out of ${n+2}$ $n$-faces in the ${(n+1)}$-simplex ${s[01\dots n+1]}$, then the other one is determined uniquely. Similarly, the conditions ${\Kan(m+1, j)}$ are satisfied for ${0\le j \le m < n}$ because the following equation, 
\begin{align}
\dd_{\alpha_m} x^m =  c_{m+1} (x^1; x^2; \dots; x^{m-1}),
\end{align}
implies that any ${m+1}$ out of the ${m+2}$ $m$-faces in the ${(m+1)}$-simplex ${s[01\dots (m+1)]}$ determines the other one. Thus, we can always fill the ${(m+1,j)}$-horn and we have unique filler if and only if ${A_{m+1}=0}$. Therefore, our simplicial set $\cB$ satisfies the required Kan conditions.

It is straightforward to identify $\cB(G)$ with one of the standard simplicial models for the classifying space $BG$ (see \Rf{GoerssJardine}, Example 1.5).
If ${n\geq 2}$, we let ${\cB(A_n,n)}$ denote the simplicial set just constructed whose data $G$, $A_k$, $\alpha_k$ and $c_{k+1}$  are all trivial except for the $n$-th abelian group $A_n$. Then ${\cB(A_n,n)}$ is equivalent to the simplicial model of an Eilenberg-MacLane space $B^nA_n$ described in \Rf{GoerssJardine}, Theorem 2.19 or Theorem 23.9 of \Rf{may}.

More generally, using Definition 8.1 and Proposition 8.2 of \Rf{may}, there is a Kan fibration
\begin{align}\label{higherGroupFibration}
\xymatrix{
\cB(A_n,n) \ar[r]^-{\iota^n} & \cB_{\al_2,c_3; \dots;
\al_n, c_{n+1}}(G, A_2, \dots, A_n)\ar[d]^-{\rho^n}  \\
&  \cB_{\al_2,c_3; \dots;
\al_{n-1}, c_{n}}(G, A_2, \dots, A_{n-1})}
\end{align}
The map $\iota^n$ is the natural inclusion: the simplices of ${\cB(A_n,n)}$ are viewed as a subset of those in ${ \cB_{\al_2,c_3; \dots;
\al_n, c_{n+1}}(G, A_2, \dots, A_n)}$ by inserting the identities of the groups ${G,A_2,..., A_{n-1}}$ for the missing labels. The map $\rho^n$ is obtained by replacing the $A_n$ labels with zeros.  Notice that the $k$-truncation defined in Definition 8.1 of \Rf{may} applied to our simplicial set ${B_{\alpha_2, c_3; \dots; \alpha_n, c_{n+1}}(G, A_2, \dots, A_n)}$ is 
\begin{equation*}
  \begin{split}
   & B^{(k)}_{\alpha_2, c_3; \dots; \alpha_n, c_{n+1}}(G, A_2, \dots, A_n) \\
 &\hspace{60pt}=  B_{\alpha_2, c_3; \dots; \alpha_k, c_{k+1}}(G, A_2, \dots, A_k)
  \end{split}
\end{equation*}
From this, we deduce inductively that the homotopy groups of $\cB^n$ are indeed given by ${\pi_1\cB^n=G}$ and ${\pi_k\cB^n=A_k}$ for ${k\geq 2}$, with the convention that ${A_k=0}$ if ${k>n}$.\footnote{In fact, the diagrams Eq.~\ref{higherGroupFibration} are part of a simplicial Postnikov, as described in \S 8 \Rf{may}.}

We see from this Kan fibration that ${\cB_{\al_2,c_3}(G,A_2)}$ is the nerve of a 2-group constructed using the $3$-cocycle
${c_3 \in Z^3(\cB(G);A_2^{\alpha_2}) \cong Z^3(G;A_2^{\alpha_2})}$. This case was discussed in~\Rf{BLm0307200}, Theorem 43, although from a different perspective.
Based on the discussion in \Rf{BLm0307200}, we see that cohomologuous choices of cocycles $c_3$  should give rise to equivalent $2$-groups, while choices of cocycles in different cohomology classes of ${H^3(G;A_2^{\alpha_2})}$ should give rise to inequivalent $2$-groups. 
We discuss this here in the context of our construction.

If two sets of canonical cochains $\{c_j\}$ and $\{{c}_j'\}$ differ by
coboundaries valued in $A_{j-1}$, we say that they are gauge equivalent. We show here that gauge equivalent cochains give rise to weakly equivalent nerves by constructing a simplicial equivalence 
\[f \colon \cB \to \cB'\] 
where ${\cB=\cB_{\alpha_2,c_3;\cdots ; \alpha_n, c_{n+1}}(G,A_2,\cdots, A_n)}$ and ${\cB'=\cB_{\alpha_2,c_3';\cdots ; \alpha_n, c_{n+1}'}(G,A_2,\cdots, A_n)}$.
We construct $f$ inductively as follows. We let ${f_0=f_1=id}$. If ${c_3-c_3'=\dd b_2}$ for
$b_2$ taking values in $A_2$, then we let ${f_2(x^2)= x^2+ b_2(x^1)}$;
further using this truncated simplicial homomorphism,  if ${f^*c_4-c_4'=\dd
b_3}$, where $b_3 $ takes values in $A_3$, then we let ${f_3(x^3)= x^3 + b_3(x^1; x^2)}$;
further using ${f_0, \dots, f_3}$, if ${f^*c_5-c_5'=\dd b_4}$ for $b_4$ taking values in
$A_4$, then we let ${f_4(x^4) = x^4 + b_4(x^1; x^2 ; x^3)}$. Proceeding inductively, we get a simplicial homomorphism $f$ defined by ${f_j \colon [\cB]_{j} \to [\cB']_{j}}$ on the $j$-simplices.

The simplicial homomorphism $f$ thus obtained is a weak equivalence since it induces an isomorphism
\[\pi_*f \colon \pi_*\cB\xrightarrow{\cong} \pi_*\cB'\]
on higher homotopy groups, which is shown by comparing the Kan fibrations \eqref{higherGroupFibration} for $\cB$ and $\cB'$. Therefore, up to weak equivalence, it suffices to specify the cohomology classes of the cocycles $c_{k+1}$.

We will need the following fact about our simplicial-set triangulations $\cB$. Given an abelian group $A$, the simplicial set triangulation $\cB$ specifies a cochain complex ${C^*(\cB,A)}$, whose cohomology is that of the realization ${K=|\cB|}$,
\begin{align}\label{cochaincomplex}
H^*( C^*(\cB;A)) \cong H^*(K;A).
\end{align}
Namely, the cochains $C^k(\cB;A)$ are the functions ${\varphi \colon [\cB]_k \to A}$. The differential ${\dd \colon C^{k-1}(\cB;A) \to C^{k}(\cB;A)}$ is given by
\[\dd\varphi(s[0\cdots k]) = \sum_{j= 0}^k (-1)^j\varphi \circ s[0\cdots \hat{j} \cdots k].\]
Further, given another simplicial set $\cB'$ with realization ${K'= |\cB'|}$ and a simplicial map ${\alpha \colon \cB \to \cB'}$, we get a homomorphism of chain complex
\[ \alpha^* \colon C^*(\cB';A) \to C^*(\cB;A) \]
defined by ${\alpha^*\varphi = \varphi\circ \alpha}$.
On cohomology, this corresponds to the homomorphism
\[|\alpha|^* \colon H^*(K';A) \to H^*(K;A)\]
on the cohomology of the realizations, where $|\alpha|$ is the realization of the map $\alpha$.

Finally, we make some remarks about the special case when the group $G$ is abelian and the homomorphisms $\alpha_k$ are all trivial. We let ${G=A_1}$ and abbreviate
\begin{equation*}
\cB = \cB_{c_3;\cdots;c_{n+1}}(A_1,A_2,\cdots, A_n).
\end{equation*}
We also use ${\cB^k= \cB_{c_3;\cdots;c_{k+1}}(A_1,A_2,\cdots, A_k)}$.
Under these triviality conditions, we can use $c_{n+1}$ to construct a simplicial homomorphism ${\kappa^{n+1} \colon \cB^{n-1} \to  \cB(A_{n}, n+1)}$ and obtain a ``homotopy fiber sequence''
\begin{equation}\label{higherGroupFibration2}
 \xymatrix{  \cB_{c_3;\cdots;c_{n+1}}(A_1,A_2,\cdots, A_{n}) \ar[d]^-{\rho^{n}} & \\
 \cB_{c_3;\cdots;c_{n}}(A_1,A_2,\cdots, A_{n-1}) \ar[r]^-{\kappa^{n+1}} & \cB(A_{n}, n+1). }
 \end{equation}
 The map $\kappa^{n+1}$ classifies the fibration $\rho^n$. Up to homotopy, $\kappa^{n+1}$ is determined by the cohomology class of $c_{n+1}$.

The map $\kappa^{n+1}$ is defined as follows. It is the constant map to the unique $d$-simplex of $ \cB(A_{n}, n+1)$ on $d$-simplices for ${0\leq d\leq n}$. Since ${[\cB(A_{n}, n+1)]_{n+1}=A_n}$, we can define ${c_{n+1} \colon [\cB^{n-1}]_{n+1} \to A_n}$. This extends to simplicial homomorphisms if and only if the action of the $\alpha$'s are all trivial and we name the resulting homomorphism $\kappa^{n+1}$.  Even if $\kappa^{n+1}$ is not a Kan fibration, it can be replaced by a Kan fibration so that the sequence Eq.~\ref{higherGroupFibration2} is a homotopy fiber sequence.

More generally (when $G$ is any group and the $\alpha$'s are allowed to be non-trivial), there also exist maps analogous to the $\kappa^{n+1}$ which classify the $\rho^n$, but their targets are classifying spaces for some twisted cohomology groups. See, e.g., Theorem 7.18 \Rf{Gitler}.

\section{Higher groups and Postnikov stages}
\label{Postnikov}

For a connected topological space $K$ (which we assume is a CW-complex, e.g., the realization of a simplicial set), there is an action of $\pi_1K$ on $\pi_nK$ for each ${n\geq 1}$. One can construct spaces $K_{\tau\leq n}$ with fibrations 
\[a^n \colon K \to K_{\tau\leq n}\]
that induce isomorphisms ${\pi_i=\pi_iK \to \pi_iK_{\tau\leq n}}$ for all ${i\leq n}$ and such that ${\pi_iK_{\tau\leq n}=0}$ for ${i>n}$. The space $K$ is weakly equivalent to the (homotopy) inverse limit
\[\xymatrix@C=0.5pc{ & & K(\pi_n,n) \ar[d]^-{i^n}   & K(\pi_{n-1},n-1) \ar[d]^-{i^{n-1}} &  & K(\pi_1,1) \ar[d]^-{i^1}_-= & \\
K \ar[r] & \cdots  \ar[r]& K_{\tau\leq n} \ar[r]^-{p^n} & K_{\tau\leq n-1} \ar[r]^-{p^{n-1}} & \cdots \ar[r] & K_{\tau\leq 1}}\]
where each $p^n$ are intermediate fibrations.
The truncations $K_{\tau\leq n}$ are called the \emph{Postnikov stages} of $K$, are homotopy $n$-types and can be viewed as the classifying spaces of the higher $n$-groups discussed above. 

We say in Appendix~\ref{hgroup} how to build the classifying space of an $n$-group using certain data. This included homomorphisms ${\alpha_k}$ which recorded the action of ${\pi_1K=G}$ on ${\pi_nK=A_n}$. We also had cocycles $c_{n+1}$ needed to inductively build our classifying spaces. We saw at the end of Appendix~\ref{hgroup} how, when $G$ is abelian and the homomorphisms $\alpha_n$ are trivial, we can use the cocycles $c_{n+1}$ to extend our fibrations of Eq.~\eqref{higherGroupFibration} to the fibration in Eq.~\eqref{higherGroupFibration2}.

This reflects a similar phenomenon in topological spaces. If $\pi_1K$ is abelian and acts trivially on $\pi_nK$ for all ${n\geq 2}$,\footnote{Under these triviality conditions, $K$ is called a \emph{simple space}. There is a generalization of the notion of $k$-invariants for non-simple spaces take value in twisted cohomology groups. 
} 
then the fibrations
\begin{equation}\label{higherGroupFibrationPost}
\xymatrix{ K(\pi_n,
n) \ar[r]^-{i^n} &   K_{\tau\leq n} \ar[d]^-{p^n}&  \\ 
&   K_{\tau\leq n-1} &
}
\end{equation}
are classified by maps ${k^{n+1} \colon K_{\tau\leq n-1} \to K(\pi_n,n+1)}$ that fit into  a homotopy fiber sequence
\begin{equation}\label{higherGroupFibrationPost2}
\xymatrix{ &   K_{\tau\leq n} \ar[d]_-{p^n}&  \\ 
&   K_{\tau\leq n-1}
\ar[r]^-{k^{n+1}} & K(\pi_n,n+1)  }
\end{equation}
Compare Eq.~\eqref{higherGroupFibrationPost} and \eqref{higherGroupFibrationPost2} to Eq.~\eqref{higherGroupFibration} and \eqref{higherGroupFibration2}. Here  $K(\pi_n,n+1)$ is a delooping of
${K(\pi_n,n)}$: 
\[ \Om K(\pi_n,n+1) \simeq K(\pi_n,n).\]
The homotopy class ${[p^n]}$
is determined by the cohomology class
\begin{align}
[k^{n+1}]\in H^{n+1}(K_{\tau \leq n-1},\pi_{n}(K)).
\end{align}
The importance of these fibrations will be made clear through our next example.

To obtain the classifying space of a higher $n$-group associated to the space $K$, we simply take $K_{\tau
\leq n}$. So, let's look at the classifying space of a higher $3$-group corresponding to ${K=S^2}$, the 2-sphere and 
looking at $S^2_{\tau \leq 3}$.  We will construct a simplicial set ${\cB_{c_4}(\Z,2;\Z,3)}$ whose realization is equivalent to $S^2_{\tau \leq 3}$. In this case,
because ${\pi_1S^2=0}$ and  ${\pi_2=\pi_3=\Z}$, we can take $S^2_{\tau\leq 1}$ to be a
point, and ${S^2_{\tau\leq 2} \simeq K(\Z,2)}$. So, the relevant portion of the
Postnikov tower is
\begin{equation}\label{eq:fibrationG}
  \xymatrix{ & S^2 \ar[d]^-{a^3} & \\
K(\Z,3) \ar[r]^-{i^3} & S^2_{\tau \leq 3} \ar[d]^-{p^3}&  \\
 &   K(\Z,2)  \ar[r]^-{k^{4}} & K(\Z,4)  }
 \end{equation}
and ${\alpha_3\colon S^2 \to S^2_{\tau \leq 3}}$ is an isomorphism on $\pi_2$ and $\pi_3$.  

We describe our simplicial-set triangulation of $S^2_{\tau \leq 3}$, 
\[\cB=\cB_{c_4}(\Z,2;\Z,3).\]
The data we use is
\begin{enumerate}
\item ${G=1}$, ${A_2=\Z}$ and ${A_3=\Z}$
\item $\alpha_2$ and $\alpha_3$ are trivial since ${G=1}$
\item ${c_3=0}$ since there are no non-trivial $3$-cocycles for the trivial group, but ${c_4 \in Z^4(\cB(\Z,2),\Z)}$ is non-trivial. We choose it to be the cup product, 
\[c_4 = x^2\smile x^2.\]
Specifically, for ${s[0\cdots 4] =(x^2_{...}) \in [\cB(\Z,2)]_4}$,
\[c_4(x^2_{...}) = x^2_{012}x^2_{234}. \]
\end{enumerate} 
Our simplicial set triangulation has one vertex and we use
$pt$ to label this vertex.  Since ${G=1}$, the triangulation also has one
link and we use ${x_{pq}^1 = 1}$ to label this unique link for any indices ${p,q}$.
We omit the links from the notation that denotes the
 triangles in ${\cB_{c_4}(\Z,2;\Z,3)}$ and simply write
\[ ( x_{01}^1, x_{12}^1, x_{02}^1;
x_{012}^2)  = (1,1,1;x_{012}^2) =(x_{012}^2) \] 
where  ${x_{012}^2 \in \pi_2(S^2) = \Z}$.  We see that the condition ${\dd x^1=1}$  is trivially satisfied, and so ${[\cB]_2=\Z}$. Geometrically, we can think of $[\cB]_2$ as a wedge of $2$-spheres labeled by the integers, where the $2$-spheres are the quotients of triangles by their boundary.

Since $c_3=0$ and $\alpha_2$ is trivial, the condition ${\dd_{\alpha_2}x^2=c_3}$ on the labels 
\begin{align}\label{eq:tetrahedronS2}
s[0\cdots3] = ( x_{012}^2 , x_{023}^2,  x_{013}^2, x_{123}^2 ; x_{0123}^3) \in \Z^4 \times \Z 
\end{align}
implies that the 3-simplices or tetrahedra in
${\cB_{c_4}(\Z,2;\Z,3)}$
satisfy
\begin{align}
\label{xxxx}
\dd x^2(s[0\cdots3])= x_{123}^2 - x_{023}^2 +x_{013}^2- x_{012}^2 =0.
\end{align}
Geometrically, this means that the 2-spheres labeled by 
\begin{align} 
(x_{012}^2), (x_{013}^2), (x_{023}^2), (x_{123}^2),
\end{align} 
together bound a tetrahedron in $\cB_{c_4}(\Z,2;\Z,3)$ when their labels satisfy Eq.~\ref{xxxx}. If this is the case, then they bound infinitely many tetrahedra, each labeled by
the additional index ${x_{0123}^3 \in \Z}$ of Eq.~\eqref{eq:tetrahedronS2}.

Note that this discussion implies the canonical 2-cochain ${x^2 \colon [\cB]_2 \to \Z}$ is actually a 2-cocycle in the cochain complex $C^*(\cB,\Z)$ of Eq.~\eqref{cochaincomplex}.

In one higher dimension, labels for 4-simplices are tuples
\begin{align}
s[0\cdots4] = (x^2_{...} ; x_{....}^3) \in \Z^{10}\times \Z^5 .
\end{align}
For such a tuple to label a 4-simplex in ${\cB_{c_4}(\Z,2;\Z,3)}$ it must first satisfy ${\dd x^2=0}$ for any of its $3$-faces or triangles, i.e.,
\[ x_{qrs}^2-x_{prs}^2 + x_{pqs}^2-x_{pqr}^2=0\]
for any ${0\leq p<q<r<s\leq 4}$. Using the canonical 3-cochain $x^3$ whose evaluation on
a tetrahedron as in Eq.~\ref{eq:tetrahedronS2} is given by $x_{0123}^3$, we also need the five tetrahedra in ${s[0\cdots4]}$ labelled by $x_{pqrs}^3$ to satisfy
\begin{align}
\label{x3x2}
 \dd x^3 = x^2\smile x^2 = c_4
\end{align}
where 
\[\dd x^3(s[0\cdots 4]) = \sum_{j=0}^4 x^3(s[0\cdots \hat{j} \cdots 4]). \]
That is,
\[ x_{012}^2x_{234}^2 = x_{1234}^3- x_{0234}^3+x_{0134}^3-x_{0124}^3+x_{0123}^3.\]

For higher dimensional simplices, there are no additional conditions: i.e., ${(x^2_{...} ; x^3_{....}) \in \Z^{\binom{d+1}{3}}\times  \Z^{\binom{d+1}{4}}}$ is a $d$-simplex in ${\cB_{c_4}(\Z,2;\Z,3)}$ if and only if all its 3-faces (tetrahedra) satisfy ${\dd x^2=0}$ and all its 4-faces satisfy ${\dd x^3= x^2\smile x^2}$.

Now we explain why the realization of our simplicial set ${\cB_{c_4}(\Z,2;\Z,3)}$ is $S^2_{\tau\leq 3}$. Eqs.~\eqref{higherGroupFibration} and \eqref{higherGroupFibration2} combine in a diagram
\[\xymatrix{
\cB(\Z;3) \ar[r]^-{\iota^3} & \cB_{c_4}(\Z,2;\Z,3) \ar[d]^-{\rho^3} & \\
& \cB(\Z;2) \ar[r]^{\kappa^4} & \cB(\Z;4)
}\]
with $\kappa^4$ constructed from $c_4$ and representing the cohomology class  
\[[c_4]  = \Sq_\Z^2 \in H^4( \cB(\Z;2);\Z) \cong H^4(K(\Z,2),\Z),\]
where $\Sq_\Z^2$ denotes the squaring operation in integral cohomology.
If we apply realization, we get a corresponding diagram 
\[\xymatrix{
K(\Z,3) \ar[r]^-{|\iota^3|} & | \cB_{c_4}(\Z,2;\Z,3)| \ar[d]^-{|\rho^3|} & \\
&  K(\Z,2)  \ar[r]^{|\kappa^4|} & K(\Z,4)
}\]
where $|\rho^3|$ is a Serre fibration 
and $|\kappa^4|$ still represents the same cohomology class.
But, as we explained in Sec.~\ref{S2Section}, $\Sq_\Z^2$ is precisely the $k$-invariant $k^4$ of the sphere $S^2$. So, 
\[ |\cB_{c_4}(\Z,2;\Z,3)|\simeq S^2_{\tau\leq 3}\]
are homotopy equivalent and we have realized Eq.~\eqref{higherGroupFibrationPost} and \eqref{higherGroupFibrationPost2} for ${K=S^2}$ when ${n=3}$.

We consider the cochain complex introduced in the discussion of Eq.~\eqref{cochaincomplex}. The simplicial maps in Eq.~\eqref{higherGroupFibration} give homomorphisms of cochain complexes
\[\xymatrix{ C^*(\cB(\Z,2);\Z) \ar[r]^-{(\rho^2)^*}   & C^*(\cB;\Z)  \ar[r]^-{(\iota^3)^*} &  C^*(\cB(\Z;3) ;\Z) }\]
where here ${\cB=\cB_{c_4}(\Z,2;\Z,3)}$ as above. We also have corresponding homomorphisms upon taking cohomology
\[\xymatrix{ H^*(K(\Z,2);\Z) \ar[r]^-{(p^2)^*}   & H^*(S^2_{\tau\leq 3};\Z) \ar[r]^-{(\iota^3)^*} &  H^*(K(\Z,3),\Z) }\]
Note we are not saying these sequences are exact, just that the maps are homomorphisms.

Consider the fundamental cochains ${y^2\in C^2(\cB(\Z,2);\Z)}$ and ${z^3 \in C^3(\cB(\Z,3);\Z)}$. We use $y$ and $z$ to avoid confusion with the fundamental cochains ${x^2\in C^2(\cB;\Z)}$ and ${x^3 \in C^3(\cB;\Z)}$. The cochains $y^2$ and $z^3$ are cocycles that represent the fundamental classes in ${H^2(K(\Z,2);\Z)\cong \Z}$ and ${H^3(K(\Z,3);\Z)\cong \Z}$. Further, we have ${(\rho^2)^*(y^2)=x^2}$ and ${(\iota^3)^*x^3=z^3}$. The first equation implies that $x^2$ is a cocycle and that it detects a generator in ${H^2(S^2_{\tau\leq 3} ;\Z) \cong \Z}$. We had already noted above that $x^2$ was a cocycle. However, even if $z^3$ is a cocycle, $x^3$ need not be one. In fact, by our construction of $\cB$, $x^3$ is not a cocycle since it satisfies Eq.~\eqref{x3x2}, i.e., ${\dd x^3= x^2 \smile x^2}$. This is consistent with the fact that ${H^3(S^2_{\tau\leq 3} ;\Z)=0}$, so that $x^3$ should not represent a cohomology class. Note that 
\begin{align*}   
\dd z^3 &=   \dd (\iota^3)^* x^3  =(\iota^3)^*   \dd x^3\\
&= (\iota^3)^* (x^2\smile x^2) = z^2 \smile z^2.
\end{align*}
But ${z^2 =0}$ in ${C^2(\cB(\Z,3);\Z)}$, so ${\dd z^3 =0}$ and $z^3$ is indeed a cocycle.

\section{Simplicial sets and Kan conditions}
\label{SSet}

In this appendix, we give some background on simplicial sets. 

\bigskip
Let $\Delta$ be the category of finite ordinals. Its objects are the ordered sets ${[n]=\{0,\cdots, n\}}$ for ${n\geq 0}$, for example,
\[
[0]=\{0\}, \quad [1]=\{0, 1\},\quad 
[2]=\{0, 1,2\},\quad\dotsc,
\]
The morphisms are the
order-preserving maps. For example, we have co-face maps
\begin{equation}\label{eq:coface}
d^i\colon [n-1] \to [n], \quad \forall j< i, j \mapsto j, \forall j \ge i ,
j \mapsto j+1,
\end{equation}
given to the inclusion that skips $i$, and co-degeneracy maps, 
\begin{equation}\label{eq:codegen}
s^i\colon [n] \to [n-1], \quad \forall j <i, j \mapsto j, \forall j\ge i,
j\mapsto j-1,
\end{equation}
given by the surjection that repeats $i$. In fact, any order-preserving
map is a composition of various $d^i$'s and $s^i$'s, and so these maps generate all the morphisms in the category $\Delta$.

Let $\Sets$ be the category of sets. A \emph{simplicial set} $X$ is a contravariant functor from the category of finite ordinals to $\Sets$, ${X\colon \Delta^{\mathrm{op}} \to \Sets}$. In other words, $X$ consists of a tower of sets ${X_0, X_1, \dots, X_n, \dots}$ with face maps ${d_i\colon X_n \to X_{n-1}}$  and degeneracy maps ${s_i\colon X_{n-1} \to X_n}$, which are dual to $d^i$ and $s^i$, i.e., ${X(d^i)=d_i}$ and ${X(s^i)=s_i}$. The category of simplicial sets is denoted $\sSets$. 

If we take a simplicial decomposition of a topological space
$K$ and take $X_n$ to be the set of $n$-simplices, then the
collection of $X_n$ form a simplicial set with $d_i$ the natural face
maps and $s_i$ the natural degeneracy maps. Thus, it is not hard to
imagine, in general, for a simplicial set $X$,  the maps $d_i$ and $s_i$
satisfy the following expected coherence conditions, 
\begin{equation}\label{eq:face-degen}
\begin{split}
 d_i d_j = d_{j-1} d_i \; \text{if}\; i<j, &\quad s_i s_j = s_{j+1} s_i \; \text{if}\; i\leq j,
 \\
 d_i s_j = s_{j-1} d_i \; \text{if}\; i<j, &\quad
 d_j s_j=\id=d_{j+1} s_j,\\
 d_i s_j = s_j d_{i-1} \; \text{if}\; i> j+1.
\end{split}
\end{equation}
 
\begin{Example} [$m$-simplex and ${(m, j)}$-horn]
If we take a geometric $n$-simplex and take its natural simplicial
decomposition, we end up with a simplicial set $\Delta^m$, which can
be described in the following combinatorial way, 
\begin{equation}
\begin{aligned} \label{eq:simplex}
 (\Simp{m})_n &= \{f\colon [n] \to [m]
  \mid f(i)\le f(j) \text{ for all }i \le j\} \\
&=  \Hom_\Delta([n],[m]).
\end{aligned} 
\end{equation}
Similarly, we define the simplicial ${(m, j)}$-horn as the
following:
\begin{equation} \label{eq:horn}
\begin{split}
(\Horn{m}{j})_n &=
  \bigl\{f\in (\Simp{m})_n\bigm| \{0,\dotsc,j-1,j+1,\dotsc,m\} \\
  &\nsubseteq \{f(0),\dotsc, f(n)\} \bigr\}.
\end{split}
\end{equation}
The geometric realization of $\Horn{m}{j}$ is an $m$-simplex with the inner and $j$-th
facet removed.  
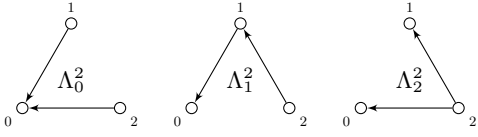
\begin{figure}[t!]
    \centering
    \begin{tikzpicture}
      [>=latex', mydot/.style={draw,circle,inner sep=1.5pt}, every label/.style={scale=0.6},scale=0.75]
      \node[scale=1] at (0,0) (a) {\(\Horn{2}{1}\)};
      \node[mydot,label=210:\(0\)]              at (-0.866,-0.5)  (a0) {};
      \node[mydot,label=90:\(1\)]    at (0,1)          (a1) {};
      \node[mydot,label=-30:\(2\)]   at (0.866,-0.5)   (a2) {};
      \path[<-]
      (a0) edge (a1) 
      (a1) edge (a2);
      \begin{scope}[xshift=-3cm,<-]
\node[scale=1] at (0,0) (a) {\(\Horn{2}{0}\)};
        \node[mydot,label=210:\(0\)]              at (-0.866,-0.5)  (b0) {};
        \node[mydot,label=90:\(1\)]    at (0,1)          (b1) {};
        \node[mydot,label=-30:\(2\)]   at (0.866,-0.5)   (b2) {};
        \path
        (b0) edge (b1)
        (b0) edge (b2);
      \end{scope}
    \begin{scope}[xshift=+3cm,<-]
\node[scale=1] at (0,0) (a) {\(\Horn{2}{2}\)};
        \node[mydot,label=210:\(0\)]              at (-0.866,-0.5)  (b0) {};
        \node[mydot,label=90:\(1\)]    at (0,1)          (b1) {};
        \node[mydot,label=-30:\(2\)]   at (0.866,-0.5)   (b2) {};
        \path
        (b1) edge (b2)
        (b0) edge (b2);
      \end{scope}
    \end{tikzpicture}
    \caption{The Horns}
    \label{fig:bimodule_A1}
  \end{figure}
Clearly, there is an inclusion of simplicial sets $
\iota_{m,j}\colon\Horn{m}{j}\to \Simp{m}$. 
\end{Example}

Then the set of simplicial morphisms ${\Hom_{\sSets}(\Simp{m}, X) = X_m}$, and
${\Hom_{\sSets}(\Horn{m}{j}, X)}$ is usually some sort of product of $X_i$'s and
represents horns in $X$. For example, ${\Hom(
\Horn{2}{1}, X) = X_1 \times_{d_0, X_0, d_1} X_1}$. 

\begin{definition}
  \label{def:Kan}
 A simplicial set $ X$ satisfies the
  Kan condition ${\Kan(m,j)}$ if and only if the
  canonical map (i.e., the horn projection)
  \begin{equation}
    \label{eq:Kan_arrow}
    X_m = \Hom_{\sSets}(\Simp{m},X)\xrightarrow{\iota^\ast_{m,j}} 
\Hom_{\sSets}(\Horn{m}{j}, X)
  \end{equation}
  is surjective. It satisfies the
  unique Kan condition ${\Kan!(m, j)}$  if and only if  the canonical map
  in \eqref{eq:Kan_arrow} is an isomorphism.
  
  We call $X$ a Kan simplicial set (or  a Kan complex or the nerve of an $\infty$-groupoid) if it satisfies ${\Kan(m,j)}$
  for all ${m\ge 1}$, ${0\le j\le m}$. We call $X$ the nerve of an $n$-groupoid\footnote{These are also called $n$-hypergroupoids in \Rf{duskin2} and $n$-groupoid (without nerve) in \cite{z:tgpd-2}.} if it satisfies ${\Kan(m,j)}$
  for all ${m\ge 1}$, ${0\le j\le m}$ and ${\Kan!(m, j)}$ for all ${m\ge
  n+1}$, ${0\le j\le m}$. Finally, $X$ is called the nerve of an $n$-group if it is the nerve of a
  $n$-groupoid with the property that $X_0$ is a point.  
\end{definition}
For the content of this Appendix, we refer to the standard textbooks~\cite{may,
  hovey} for the theory simplicial sets. $\infty$-groupoids using Kan
condition are due to~\cite{duskin2}, we also refer to the 
in \cite[Sect.1]{z:tgpd-2} for a nice detailed introduction of this topic.

\section{An informal calculation of ${c_4}$}\label{QuickPostnikov}

This appendix presents a quick, informal calculation of Eq.~\eqref{c4forS2} of the main text. Consider the ${\mathbb{C}P^1}$ presentation of the ${S^2}$ nonlinear ${\si}$-model. As mentioned in Sec.~\ref{S2Section}, in this presentation, the homotopy classes ${\pi_2(S^2) \simeq \Z}$ are characterized by the Chern number of the $U(1)$-bundle,
\begin{align}\label{s2App}
 \int_{S^2} \frac{1}{2\pi} \dd A \in \pi_2(S^2)\simeq\Z,
\end{align}
while the homotopy classes ${\pi_3(S^2)}$ are characterized by the Hopf invariant, which can be expressed in terms of $A_\mu$ as~\cite{Ah0005150}
\begin{align}\label{s3App}
 \int_{S^3} \frac{1}{4\pi^2}  A\wdg\dd A \in \pi_3(S^2)\simeq\Z.
\end{align}

To determine whether or not the Postnikov ${4}$-invariant ${c_4}$ is nontrivial, we introduce the maps
\begin{equation}
\phi_{k}\colon S^k \to \cB_{c_4}( 0,\Z,\Z),
\end{equation}
where ${S^k}$ is a ${k}$-sphere and ${k=1,2,3}$. In terms of the ${n}$-cochains ${x^n}$ on ${\cB_{c_4}( 0,\Z,\Z)}$, the homotopy classes of these maps are characterized by
\begin{equation}\label{xkApp}
\int_{S^k}\phi^*_{k}~x^k \in \pi_k(S^2).
\end{equation}

Comparing Eq.~\eqref{xkApp} to Eqs.~\eqref{s2App} and~\eqref{s3App}, we relate the cochains ${x^k}$ to the ${U(1)}$ gauge field ${A}$ by
\begin{align}
 \phi_2^* x^2 &= \frac{1}{2\pi} \dd A,\\
 \phi_3^* x^3 &= \frac{1}{4\pi^2}  A \wdg\dd A + \text{coboundary}.
\end{align}
These imply that ${x^2}$ and ${x^3}$ satisfy
\begin{equation}
\dd x^2 = 0,\quad\quad \dd x^3 = x^2\smile x^2,
\end{equation}
Comparing this to Eq.~\eqref{S23grpcocycleconds} in the main text, we find that 
\begin{equation}
c_4(x^2) = x^2\smile x^2,
\end{equation}
as claimed by Eq.~\eqref{c4forS2}.

\let\oldaddcontentsline\addcontentsline
\renewcommand{\addcontentsline}[3]{}
\bibliography{all,allnew,publst,publstnew,local,salRefs}
\let\addcontentsline\oldaddcontentsline

\end{document}